\documentclass{aastex61}

\usepackage{bm}
\usepackage{amsmath}

\catcode`\@=11
\newcommand{\gsim}{${\mathrel{\mathpalette\@versim>}}$}
\newcommand{\lsim}{${\mathrel{\mathpalette\@versim<}}$}
\newcommand{\@versim}[2]{\lower 2.9truept \vbox{\baselineskip 0pt \lineskip
    0.5truept \ialign{$\m@th#1\hfil##\hfil$\crcr#2\crcr\sim\crcr}}}
\catcode`\@=12

\newcommand{\ddeg}{$^{o}$}
\newcommand\HI{H\,{\sc i}}
\newcommand\HII{H\,{\sc ii}}

\def\be{\begin{equation}}
\def\bea{\begin{eqnarray}}
\def\ee{\end{equation}}
\def\eea{\end{eqnarray}}

\received{}
\revised{}
\accepted{}
\submitjournal{ApJ}

\shorttitle{Phased Array Feed}
\shortauthors{Roshi et al.}

\begin{document}

\title{
Performance of a highly sensitive, 19-element, dual-polarization, \\ cryogenic
L-band Phased Array Feed on the \\ Green Bank Telescope
}

\correspondingauthor{D. Anish Roshi, W. Shillue}
\email{aroshi@nrao.edu, bshillue@nrao.edu}

\author[0000-0002-0786-7307]{D. Anish Roshi}
\affil{National Radio Astronomy Observatory (NRAO)\footnote{
The National Radio Astronomy Observatory is a facility of
the National Science Foundation operated under a cooperative
agreement by Associated Universities, Inc.}, \\
520 Edgemont Road, Charlottesville, VA 22903, USA}

\author{W. Shillue}
\affil{National Radio Astronomy Observatory (NRAO)\footnote{
The National Radio Astronomy Observatory is a facility of
the National Science Foundation operated under a cooperative
agreement by Associated Universities, Inc.}, \\
520 Edgemont Road, Charlottesville, VA 22903, USA}

\author{B. Simon}
\affiliation{Green Bank Observatory (GBO)\footnote{
The Green Bank Observatory is a facility of the National Science Foundation
operated under a cooperative agreement by Associated Universities, Inc.},
155 Observatory Rd, \\ Green Bank, WV 24944, USA}

\author{K. F. Warnick}
\affiliation{Brigham Young University (BYU), Provo, UT 84602, USA}

\author{B. Jeffs}
\affiliation{Brigham Young University (BYU), Provo, UT 84602, USA}

\author{D. J. Pisano}
\affiliation{West Virginia University (WVU), Dept. of Physics \& Astronomy, \\
P.O. Box 6315, Morgantown, WV 26506, USA.}
\affiliation{Center for Gravitational Waves and Cosmology, WVU, \\ Chestnut Ridge Research Building,
Morgantown, WV 26505, USA.}

\author{R. Prestage}
\affiliation{Green Bank Observatory (GBO)\footnote{
The Green Bank Observatory is a facility of the National Science Foundation
operated under a cooperative agreement by Associated Universities, Inc.},
155 Observatory Rd, \\ Green Bank, WV 24944, USA}

\author{S. White}
\affiliation{Green Bank Observatory (GBO)\footnote{
The Green Bank Observatory is a facility of the National Science Foundation
operated under a cooperative agreement by Associated Universities, Inc.},
155 Observatory Rd, \\ Green Bank, WV 24944, USA}

\author{J. R. Fisher}
\affil{National Radio Astronomy Observatory (NRAO)\footnote{
The National Radio Astronomy Observatory is a facility of
the National Science Foundation operated under a cooperative
agreement by Associated Universities, Inc.}, \\
520 Edgemont Road, Charlottesville, VA 22903, USA}

\author{M. Morgan}
\affil{National Radio Astronomy Observatory (NRAO)\footnote{
The National Radio Astronomy Observatory is a facility of
the National Science Foundation operated under a cooperative
agreement by Associated Universities, Inc.}, \\
520 Edgemont Road, Charlottesville, VA 22903, USA}

\author{R. Black}
\affiliation{Brigham Young University (BYU), Provo, UT 84602, USA}

\author{M. Burnett}
\affiliation{Brigham Young University (BYU), Provo, UT 84602, USA}

\author{J. Diao}
\affiliation{Brigham Young University (BYU), Provo, UT 84602, USA}

\author{M. Ruzindana}
\affiliation{Brigham Young University (BYU), Provo, UT 84602, USA}

\author{V. van Tonder}
\affiliation{Green Bank Observatory (GBO)\footnote{
The Green Bank Observatory is a facility of the National Science Foundation
operated under a cooperative agreement by Associated Universities, Inc.},
155 Observatory Rd, \\ Green Bank, WV 24944, USA}
\affiliation{Square Kilometre Array South Africa (SKA SA), Cape Town, South Africa.}

\author{L. Hawkins}
\affiliation{Green Bank Observatory (GBO)\footnote{
The Green Bank Observatory is a facility of the National Science Foundation
operated under a cooperative agreement by Associated Universities, Inc.},
155 Observatory Rd, \\ Green Bank, WV 24944, USA}

\author{P. Marganian}
\affiliation{Green Bank Observatory (GBO)\footnote{
The Green Bank Observatory is a facility of the National Science Foundation
operated under a cooperative agreement by Associated Universities, Inc.},
155 Observatory Rd, \\ Green Bank, WV 24944, USA}

\author{T. Chamberlin}
\affiliation{Green Bank Observatory (GBO)\footnote{
The Green Bank Observatory is a facility of the National Science Foundation
operated under a cooperative agreement by Associated Universities, Inc.},
155 Observatory Rd, \\ Green Bank, WV 24944, USA}

\author{J. Ray}
\affiliation{Green Bank Observatory (GBO)\footnote{
The Green Bank Observatory is a facility of the National Science Foundation
operated under a cooperative agreement by Associated Universities, Inc.},
155 Observatory Rd, \\ Green Bank, WV 24944, USA}

\author{N. M. Pingel}
\affiliation{West Virginia University (WVU), Dept. of Physics \& Astronomy, \\
P.O. Box 6315, Morgantown, WV 26506, USA.}
\affiliation{Center for Gravitational Waves and Cosmology, WVU, \\ Chestnut Ridge Research Building,
Morgantown, WV 26505, USA.}

\author{K. Rajwade}
\affiliation{West Virginia University (WVU), Dept. of Physics \& Astronomy, \\
P.O. Box 6315, Morgantown, WV 26506, USA.}
\affiliation{Center for Gravitational Waves and Cosmology, WVU, \\ Chestnut Ridge Research Building,
Morgantown, WV 26505, USA.}

\author{D.R. Lorimer}
\affiliation{West Virginia University (WVU), Dept. of Physics \& Astronomy, \\
P.O. Box 6315, Morgantown, WV 26506, USA.}
\affiliation{Center for Gravitational Waves and Cosmology, WVU, \\ Chestnut Ridge Research Building,
Morgantown, WV 26505, USA.}

\author{A. Rane}
\affiliation{West Virginia University (WVU), Dept. of Physics \& Astronomy, \\
P.O. Box 6315, Morgantown, WV 26506, USA.}
\affiliation{Center for Gravitational Waves and Cosmology, WVU, \\ Chestnut Ridge Research Building,
Morgantown, WV 26505, USA.}

\author{J. Castro}
\affil{National Radio Astronomy Observatory (NRAO)\footnote{
The National Radio Astronomy Observatory is a facility of
the National Science Foundation operated under a cooperative
agreement by Associated Universities, Inc.}, \\
520 Edgemont Road, Charlottesville, VA 22903, USA}

\author{W. Groves}
\affil{National Radio Astronomy Observatory (NRAO)\footnote{
The National Radio Astronomy Observatory is a facility of
the National Science Foundation operated under a cooperative
agreement by Associated Universities, Inc.}, \\
520 Edgemont Road, Charlottesville, VA 22903, USA}

\author{L. Jensen}
\affiliation{Green Bank Observatory (GBO)\footnote{
The Green Bank Observatory is a facility of the National Science Foundation
operated under a cooperative agreement by Associated Universities, Inc.},
155 Observatory Rd, \\ Green Bank, WV 24944, USA}

\author{J. D. Nelson}
\affiliation{Green Bank Observatory (GBO)\footnote{
The Green Bank Observatory is a facility of the National Science Foundation
operated under a cooperative agreement by Associated Universities, Inc.},
155 Observatory Rd, \\ Green Bank, WV 24944, USA}

\author{T. Boyd}
\affil{National Radio Astronomy Observatory (NRAO)\footnote{
The National Radio Astronomy Observatory is a facility of
the National Science Foundation operated under a cooperative
agreement by Associated Universities, Inc.}, \\
520 Edgemont Road, Charlottesville, VA 22903, USA}

\author{A.J. Beasley}
\affil{National Radio Astronomy Observatory (NRAO)\footnote{
The National Radio Astronomy Observatory is a facility of
the National Science Foundation operated under a cooperative
agreement by Associated Universities, Inc.}, \\
520 Edgemont Road, Charlottesville, VA 22903, USA}

\begin{abstract}
  A new 1.4~GHz 19-element, dual-polarization, cryogenic phased array
  feed (PAF) radio astronomy receiver has been developed for the
  Robert C. Byrd Green Bank Telescope (GBT) as part of FLAG (Focal
  L-band Array for the GBT) project.  Commissioning observations of
  calibrator radio sources show that this receiver has the lowest
  reported beamformed system temperature ($T_{\rm sys}$) normalized by
  aperture efficiency ($\eta$) of any phased array receiver to
  date. The measured $T_{\rm sys}/\eta$ is $25.4 \pm 2.5$~K near 1350~MHz
  for the boresight beam, which is comparable to the performance of
  the current 1.4~GHz cryogenic single feed receiver on the GBT. The
  degradation in $T_{\rm sys}/\eta$ at $\sim$ 4\arcmin\
  (required for Nyquist sampling) and $\sim$ 8\arcmin\ offsets from
  the boresight is, respectively, $\sim$~1\% and $\sim$~20\% of
  the boresight value. The survey speed of the PAF with seven formed beams
  is larger by a factor between 2.1 and 7 compared to a single beam system depending
  on the observing application.
  The measured performance,
  both in frequency and offset from boresight, qualitatively agree
  with predictions from a rigorous electromagnetic model of the
  PAF. The astronomical utility of the receiver is demonstrated
  by observations of the pulsar B0329+54 and an extended
  \HII\ region, the Rosette Nebula. The enhanced survey speed with the
  new PAF receiver will enable the GBT to carry out exciting new
  science, such as more efficient observations of diffuse, extended
  neutral hydrogen emission from galactic in-flows and searches for
  Fast Radio Bursts.
\end{abstract}


\section{Introduction}

Radio Astronomy is primarily an observational science, in which
high-sensitivity, large-scale surveys are an essential tool for new
discoveries \citep{condonetal1998,barnesetal2001,manchester2001,
bekhtietal2016,dezotti2010}.
While large apertures enhance sensitivity they have a more limited
field-of-view (FoV), when equipped with a single feed.
The survey speed of a telescope
is proportional to its FoV and the square of the signal-to-noise ratio
required for the survey. In this paper, we describe a new 1.4~GHz (L-band) Phased Array Feed (PAF)
system, built for the Robert C. Byrd Green Bank Telescope (GBT) as
part of FLAG (Focal L-band Array for the GBT) project, which
can enhance the FoV of the telescope by a factor of seven and has
sensitivity comparable to the existing single feed 1.4~GHz receiver.

There are wide-ranging scientific interests to increase the FoV of the
GBT near L-band.  Here we focus on two scientific motivations. It is now
known that galaxies must be accreting gas to sustain their star
formation. With the current known gas content of galaxies, current
star formation rates can be sustained only for a few billion
years. However, there is evidence that the gas content of galaxies has
remained roughly constant for the past 10 billion years, implying that
galaxies must be accreting gas from the intergalactic medium
\citep{prochaskaetal2005}. Current theories suggest that this
accretion occurs in one of two modes: a hot mode where the gas is
heated to 10$^6$~K, or a cold mode where gas remains below 10$^5$~K
\citep{kereetal2005,kereetal2009, sancisietal2008}. The cold mode, in
particular, should be the dominant form of accretion for low mass
galaxies in low density environments.  Such accretion should be
detectable via observations of neutral hydrogen at 21-cm (HI) at
column densities of N(\HI) \lsim $10^{18}$~cm$^{2}$
\citep{poppingetal2009}. To detect \HI\ emission from the cold
accretion flows from nearby galaxies, sensitive single-dish
observations are essential \citep{chynowethetal2008, mihosetal2012, wolfeetal2013,
pisano2014, debloketal2014, wolfeetal2016}. Since the expected
line strengths are weak such surveys require enormous amounts of observing
time. To reduce the observing time it is essential to increase the FoV
of the telescope with high sensitivity multi-beam systems.

The discovery of Fast Radio Bursts (FRBs) has generated renewed
interest in exploring the transient radio sky \citep{lorimeretal2007,
katz2016}. FRBs are bright, approximately millisecond duration individual
pulses from compact sources exhibiting pulse dispersive delays over a
wide bandwidth. The dispersive delays far exceed values expected from
the interstellar medium of our Galaxy, indicating that they are
extragalactic in origin. This has been confirmed by the localization
and optical counterpart identification for FRB~121103, the only known
repeating burst source \citep{chatterjee2017direct}. So far about 24
FRBs have been detected \citep{petroffetal2016}, with most of the
detections made near 1.4~GHz. The physical nature and mechanism
producing the burst are not known. Studies of such short duration
radio transients are poised for a revolution with large FoV
telescopes and flexible, high throughput backends
\citep[see, for example,][]{bannisteretal2017}.

The FoV of a radio telescope when equipped with a single feed is
limited to the full width at half maximum (FWHM) of the primary beam,
which is $\sim \lambda/D$, where $\lambda$
is the observing wavelength and $D$ is the aperture diameter of the
reflector. The feeds are usually optimized to maximally receive
radiation from the reflector while attenuate the ground spillover. The
physical size of the feed is determined by this optimization process.
Traditionally, the FoV of telescopes is increased by placing multiple
such optimized feeds in the focal plane -- referred to as a focal
plane array (FPA). The physical separation between feeds in an FPA is
determined by the size of the feeds. The large physical separation of
the feeds results in non-overlapping beams on the sky. Further, the
off-axis beams suffer efficiency degradation since the feed
optimization is usually done for the central beam. Both of these
effects result in a relatively poor mapping efficiency
improvement.

In the last decade, the use of phased arrays employing a set of
smaller focal plane radiating elements -- referred to as phased-array
feeds (PAFs)
-- has gained wide interest among the radio astronomy community
\citep{fisherbradley2000, haybird2015,warnicketal2016}.
In such PAF receivers, each element is
electrically small, and thus does not optimally illuminate the
reflector. However an optimal illumination with low spillover is
obtained by the weighted sum of the amplified signal voltages from
multiple elements.  Multiple beams can be formed by adding signals
with different sets of complex weights.  The beams formed in this
manner can be made to overlap, thus increasing the mapping
efficiency. The design and optimization of PAFs are
complicated by mutual coupling effects between elements \citep[see, for example,][]{diaowarnick2017}.
The mutual
coupling distorts the element beam pattern, and introduces
channel-to-channel noise coupling between neighboring
low-noise-amplifiers (LNAs) which follow these elements. Thus, accurate
methods for electromagnetic modeling and beamforming are needed in
order to achieve efficient performance.

Currently, PAFs are being developed for both single-dish radio
telescopes and radio interferometers.  Designing efficient PAFs
require proper accounting of electromagnetic coupling and beamforming,
which has led to new theories and techniques in PAF modeling
\citep{woestenburg2005, warnickjensen2007, hay2010}. Further, several
leading radio astronomy institutions have made significant progress on
developing PAF receivers and systems.  The Westerbork Synthesis Radio
Telescope (WSRT) \citep{oosterlooetal2010} and the Australian Square
Kilometer Array Prototype (ASKAP) \citep{hayosullivan2008, chippendaleetal2015} have
both built uncooled, broadband ($>800$~MHz) PAFs near 1.4~GHz. The
elements used by WSRT for their PAF, called the APERture Tile In Focus
(APERTIF), are Vivaldi antennas.  The ASKAP Chequerboard PAF is made
of a self-complementary connected element array
\citep{hayosullivan2008}.  A PHased Array feed Demonstrator (PHAD)
has been developed by the Dominion Radio Astrophysical Observatory
(DRAO) using Vivaldi elements \citep{veidtetal2011}. Cryogenic PAF
development is being pursued for the Arecibo telescope by the Cornell
University \citep{cortesmedellin2015} and for the Five hundred meter
Aperture Spherical Telescope (FAST) \citep{wuetal2016}; both have
operating frequencies near 1.4~GHz. A higher frequency (70 -- 95~GHz)
PAF is also being developed by the University of Massachusetts
\citep{ericksonetal2015}.

FLAG is a collaborative project between the National Radio Astronomy
Observatory (NRAO), the Green Bank Observatory (GBO), Brigham Young
University (BYU) and West Virginia University (WVU). During the first
phase of the project, a prototype cryogenic `kite dipole array' was
built and tested successfully on the GBT \citep{roshietal2015}.
In this paper, we present the construction of the next generation cryogenic PAF,
and describe the measurement of its performance and comparison with a PAF model.
The new PAF is optimized for the GBT
optics to provide the required FoV, with a diameter of $\sim$
20\arcmin . Additionally, all of the instrumentation from the LNA to
the backend was upgraded.  A brief description of the new system is
given in Section~\ref{secinst}. For testing and commissioning of the
system on the GBT, Nyquist-sampled voltage data were recorded in an
instantaneous bandwidth of 300~kHz, and all signal processing was done
off-line. The data processing method is described in
Section~\ref{secpm_dp}.  Before the system was installed on the GBT,
the receiver temperature of the PAF was measured at the outdoor test
facility at the GBO, as described in Section~\ref{secrec}. The test
observations made with the GBT are summarized in Section~\ref{secobs}
and the results are given in Section~\ref{result}.  We have also
observed a pulsar and an extended continuum source using the
PAF system on the GBT. The results of these observations are presented in Section~\ref{astron}.
The main results are summarized in
Section~\ref{sumcon}.  A GPU-based digital signal processing backend
instrument has been realized for real-time calibration and beamforming
by FLAG collaboration. This system will form the backend for the
PAF for science observations with the GBT.  The details of the
real-time beamformer will be published elsewhere.

\section{Instrumentation}
\label{secinst}

\begin{figure}[!t]
\plotone{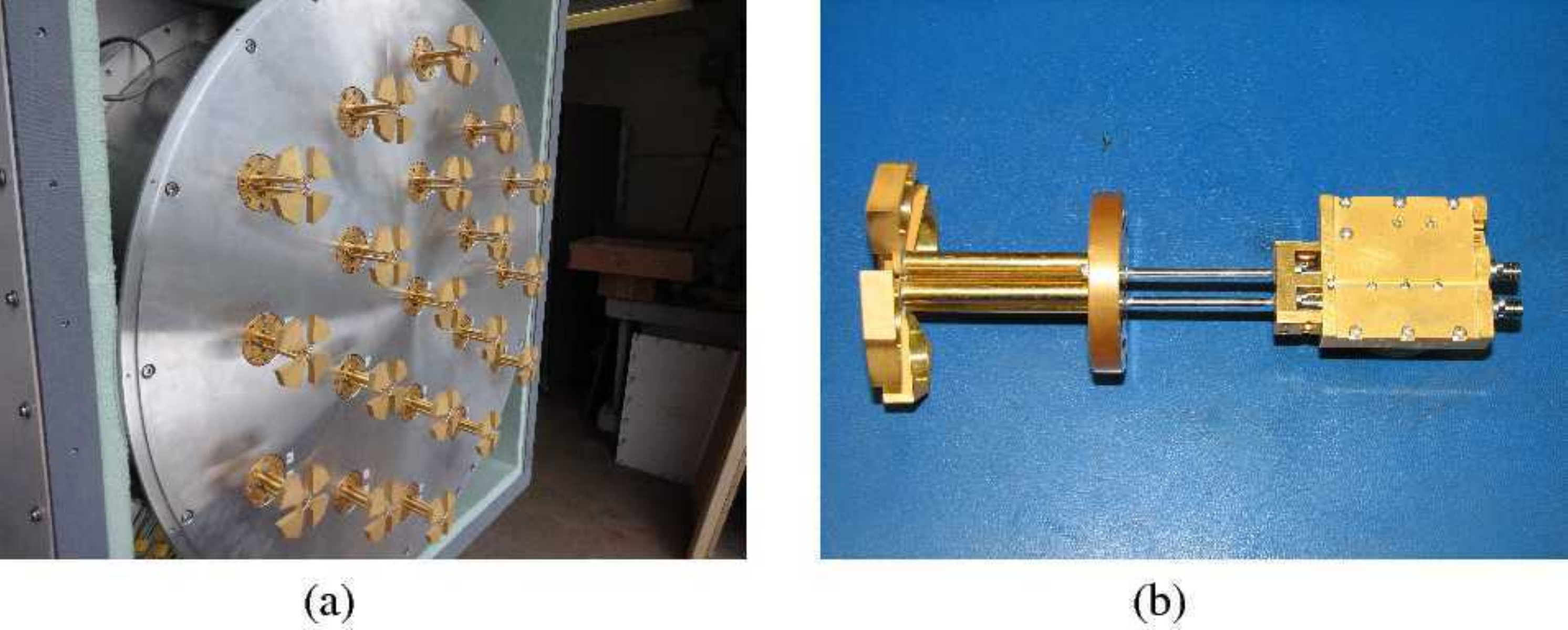}
\caption{(a) The 19 element dual-polarized dipole array.
(b) The dipole and custom low-loss, low thermal conductivity transition to the
low-noise amplifier.
\label{fig1}
}
\end{figure}

\begin{figure}[!t]
\plotone{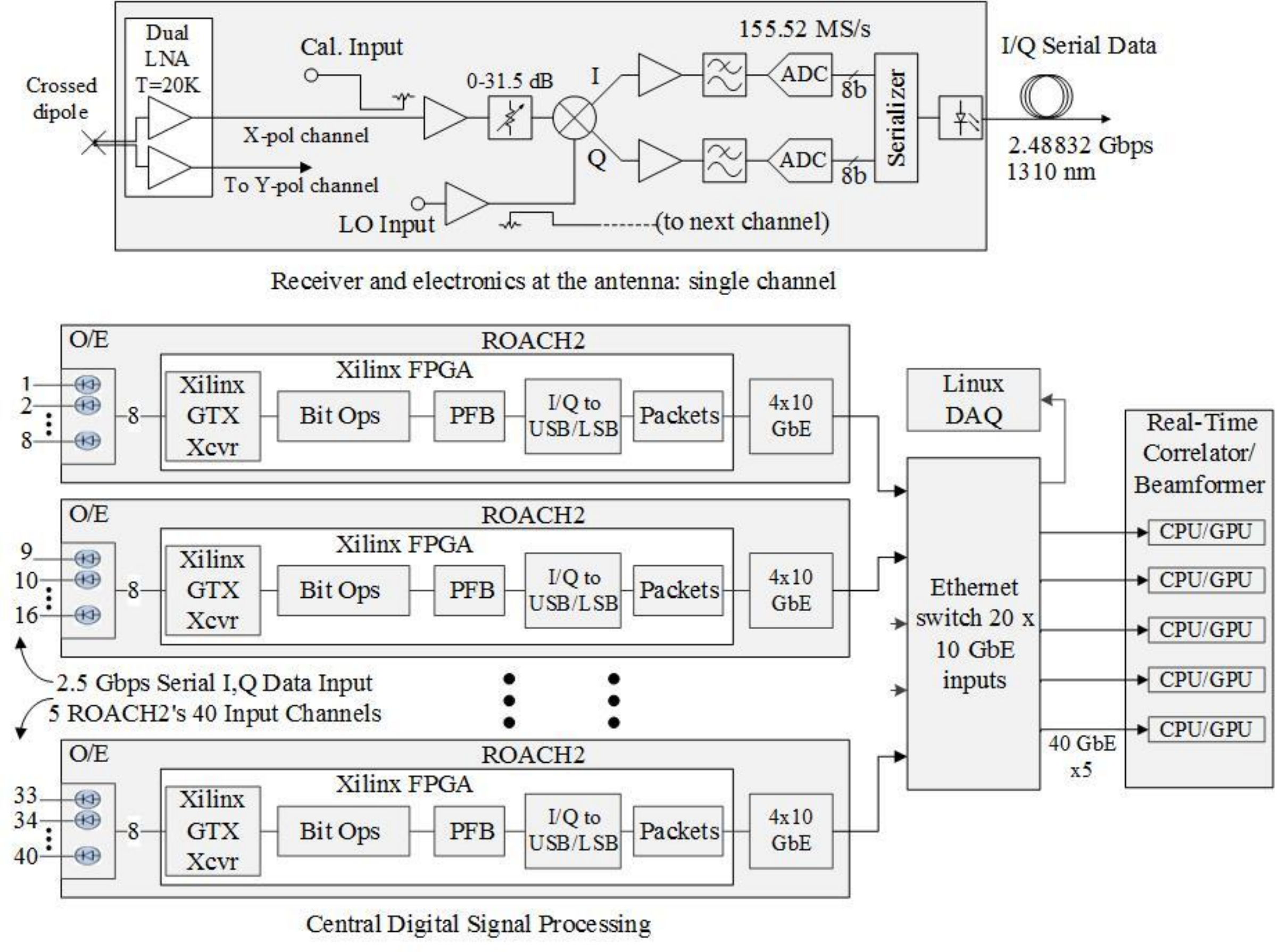}
\caption{A block diagram showing the design of the phased array receiver system.
\label{fig2}
}
\end{figure}

The PAF consists of a hexagonal array of 19 dual-polarization dipole
elements at ambient temperature connected to 19 pairs of cryogenic
LNAs located in a vacuum dewar. Optimization of the design for maximum
sensitivity over the antenna FoV of angular diameter $\sim$~20\arcmin,
and across a bandwidth of 150~MHz, required adjusting the geometric
parameters of the dipoles which had been developed in prior work
\citep{warnicketal2011}. The optimal design is dependent upon the
noise parameters of the LNA and uses maximum signal-to-noise ratio
(S/N) beamforming as a parameter. The dipole array is shown in
Fig.~\ref{fig1}a.  The dipoles were fabricated using brass and
subsequently plated with copper (with thickness $\approx 6.35 \mu$m) followed by
1~$\mu$m of gold to reduce ohmic loss. An air-filled coaxial conductor
transmission line was used to minimize dielectric loss. These vertical
transmission lines function as a balun and a means of fixing the
correct separation between the radiating elements and the ground
plane. Two teflon beads located at either end of the transmission line
center the inner conductor. Between the dipoles and the LNAs, which
are located in the cryostat, there is a custom low-loss coaxial
assembly that also serves as a thermal transition and a vacuum
barrier, as shown in Fig.~\ref{fig1}b.  The LNAs use SiGe transistors
selected for their low-noise performance at cryogenic temperatures
\citep{weinrebetal2009}.  The measured average gain of the LNAs is 38~dB
over their 1.2--1.7~GHz frequency range, and the median noise
temperature is 4.85~K (the peak-to-peak variation in the measured
values is 1.5~K), when operated at a physical temperature of 15~K
\citep{grovesmorgan2017}. A highly integrated, 40-channel electronics
assembly encompasses all of the remaining functions of the receiver:
calibration signal injection, warm post-amplification, power leveling,
local oscillator distribution, down conversion, analog-to-digital
conversion, and serial data transmission through optical fiber, as
shown in Fig.~\ref{fig2}. This post-amplification electronics assembly
is the first deployment of a novel integrated unformatted digital link
developed at NRAO \citep{morganetal2013}, which allowed the entire
analog signal path and digitizers to be located directly behind the
phased array. This reduced the power dissipation and minimized the
physical footprint. We have confirmed many aspects of this new design
including the methodology for recovering bit and word boundaries in
the unformatted received data streams \citep{morganefisher2014}. The
digital link can be scaled in bit rate, bandwidth, and channel count
for future PAF applications.  The fiber optic link transports the
samples over approximately 2~km to the Jansky Laboratory, to terminate
in equipment racks containing five ROACH2 FPGA
boards\footnote{\url{https://casper.berkeley.edu/wiki/ROACH-2_Revision_2}},
which perform bit and byte alignment, 512 channel polyphase filter
bank, and sideband separation operations. The data from the
spectral channels are re-quantized to eight bits, packetized
and sent to an Ethernet switch through 10~GbE links. Only 500
spectral channels, selected by removing channels from the band edge,
are sent to the switch. The data acquisition system developed for
commissioning and testing consists of a Linux computer equipped with a
high-speed disk, and connected to the switch through a 10~GbE
link. The complex voltage samples from one of the spectral channels is
recorded to the disk for offline processing.

The PAF and the associated receiver box are placed at the prime focus
of the GBT. The GBT local oscillator (LO) system is used for the first
down conversion. The digital down converters and the backend system
are synchronized through a 10~MHz observatory-wide frequency
reference, which is also used to lock the GBT LO system.  The LO
frequencies were set to 1550, 1450, 1350 or 1250~MHz for most
observations since the sideband separation coefficients were
previously calibrated for these LO values. For each LO setting, data
were collected from a subset of 300~kHz PFB channels. The RF frequency
corresponding to this subset of PFB channels was located within the
150~MHz bandwidth centered at the LO value.

For future science observations, a complete GPU-based digital signal
processing back end instrument has been realized to process all 500
channels from the FPGA boards, covering 150~MHz bandwidth (see
Fig.~\ref{fig2}). It performs phased array calibration, beamforming,
correlation, and pulsar and transient searches. It has several unique
capabilities not found on other PAF-equipped telescopes. Array
covariance matrices are saved continuously from the correlator as the
primary data product. This enables post-correlation beamforming \citep{haybird2015} after
the fact for spectral line or continuum observations, so that
different beam forming weights can be applied to improve beam
patterns, or to compute any desired number of overlapping beams on the
sky (within the limits of the FoV accessible to the finite number of
PAF elements). Also, we are in the process of developing software for
two new operational modes: (a) a commensal mode, where a quick-dump
real-time beamformed spectrometer runs concurrently with the
correlator to permit opportunistic transient searches during
observations of neutral hydrogen; (b) an RFI mitigation mode, where
the correlator and real-time beamformer can be linked to form a
tracking RFI nulling adaptive beamformer. This instrument was
commissioned separately on the GBT in summer 2017, and the results
will be presented elsewhere.

\section{Data Processing and Performance Metric}
\label{secpm_dp}

The offline processing of the recorded data proceeded with first
taking a 64-point Fourier transform of the complex voltage time
series, which provided a spectral resolution of $\sim 300/64 = 4.7$~kHz.
The cross correlations between signals from all the 38 dipole
outputs for each spectral channel were then computed and the
correlation matrix $\bm R$ was obtained as
\be
\bm R = \frac{1}{N} \sum_{i=0}^{N-1} \bm V[i] \bm V[i]^H.
\ee
Here $\bm V[i]$ is the time series of complex voltage vector formed from the 38 dipole outputs
from a spectral channel.
The number of samples $N$ used for the computation typically
corresponds to an integration time of 5~s.
The high spectral resolution is very useful to excise narrow
band RFI. After RFI editing the cross correlations were averaged over
300~kHz bandwidth.

An offline  post-correlation beamformer is implemented in a MATLAB
\footnote{\url{https://www.mathworks.com/products/matlab.html}} program \citep{jeffsetal2008, haybird2015}.
For typical observations, the cross correlations on
the source and at a nearby off-source position were measured. In this case,
\be
\mbox{S/N}  =  \frac{\bm w^H \bm R_{\rm on} \bm w }{\bm w^H \bm R_{\rm off}
\bm w} - 1,
\label{snr}
\ee
where $\bm R_{\rm on}$ and $\bm R_{\rm off}$ represent the on-source
and off-source correlation matrices respectively, and $\bm w$ is the
beamformer weight. The beamformer weights are obtained by maximizing
the S/N, which will be the eigenvector corresponding to the maximum
Rayleigh quotient.  For forming beams at different positions in the
FoV of the PAF system, the above procedure is repeated by moving the
telescope and positioning the source appropriately.

We define the performance metric as the inverse of the maximum S/N
when observing a compact astronomical source \citep{warnickjeffs2008}.
As shown below, this inverse S/N can be expressed in terms of $T_{\rm
  sys}/\eta$, if the flux density of the source is known, and it can
be directly obtained from the observations of a compact source without
making any assumptions. This metric is useful to compare the
performance of different PAFs. A nearby off-source position needs to
be observed to derive $T_{\rm sys}/\eta$ and hence its value will
depend on the diffuse foreground emission temperature at the
off-source position.  The measured value of $\mbox{S/N} =
\frac{T_a}{T_{\rm sys}}$, for the off-course system temperature
$T_{\rm sys}$ and the excess antenna temperature
\be
T_a = \frac{S A \eta}{2\,k}.
\label{ta}
\ee
Here $S$ is the flux density of the source, $A$ is the physical area
of the telescope aperture, $k$ is Boltzmann's constant and $\eta$ is
the product of the aperture efficiency of the telescope and
radiation efficiency of the PAF. Substituting in \eqref{ta}),
we get:
\be
\frac{T_{\rm sys}}{\eta} = \frac{S A}{2\; k \; \mbox{S/N}_{\rm max}},
\label{tsyseta}
\ee
where S/N$_{\rm max}$ is the maximum S/N. We use the flux density of
calibrator sources provided by \cite{perleybutler2017}. These
flux densities are accurate to 3--5\%.

\section{Receiver Temperature}
\label{secrec}

\begin{figure}[!t]
\plottwo{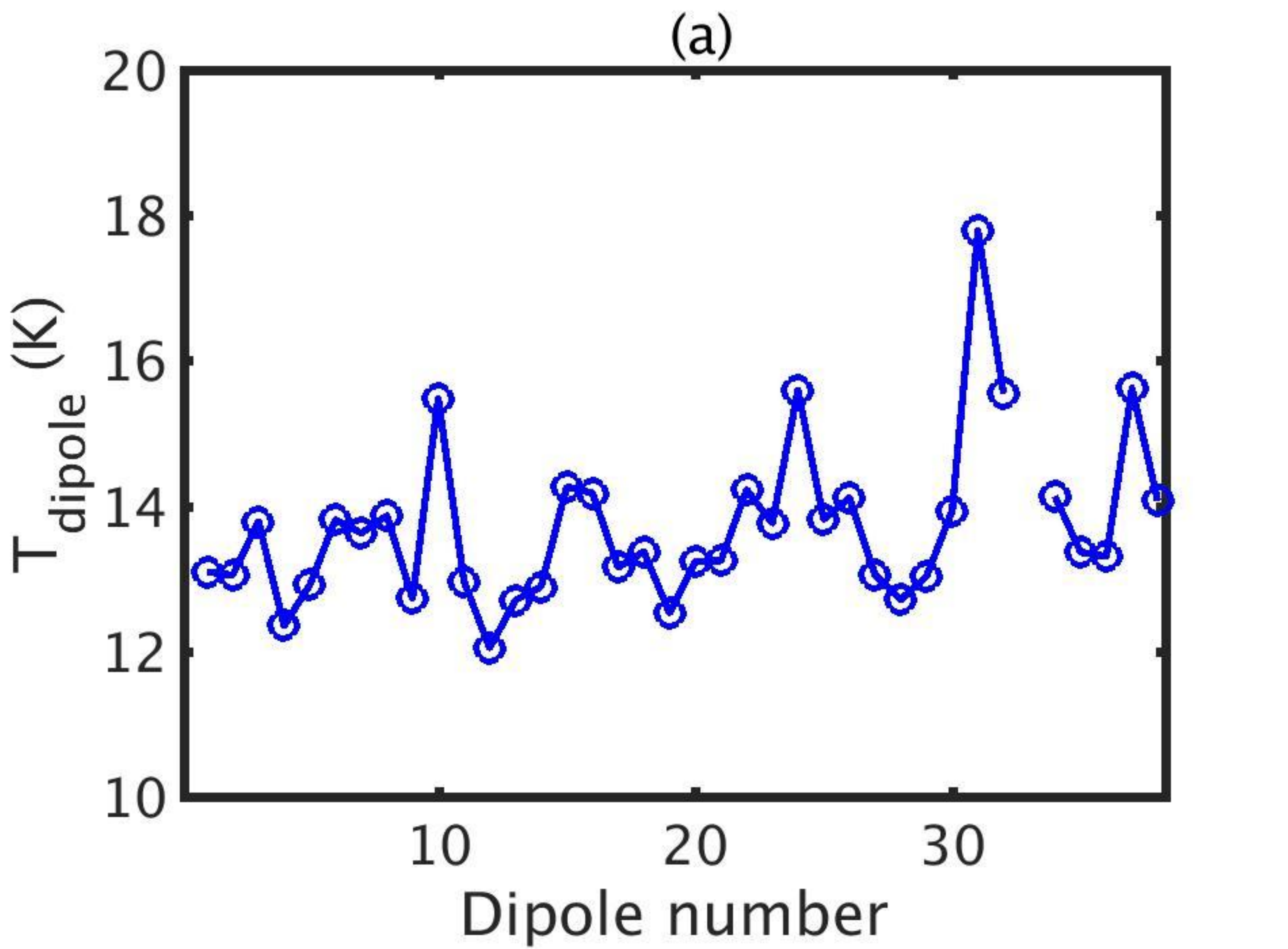}{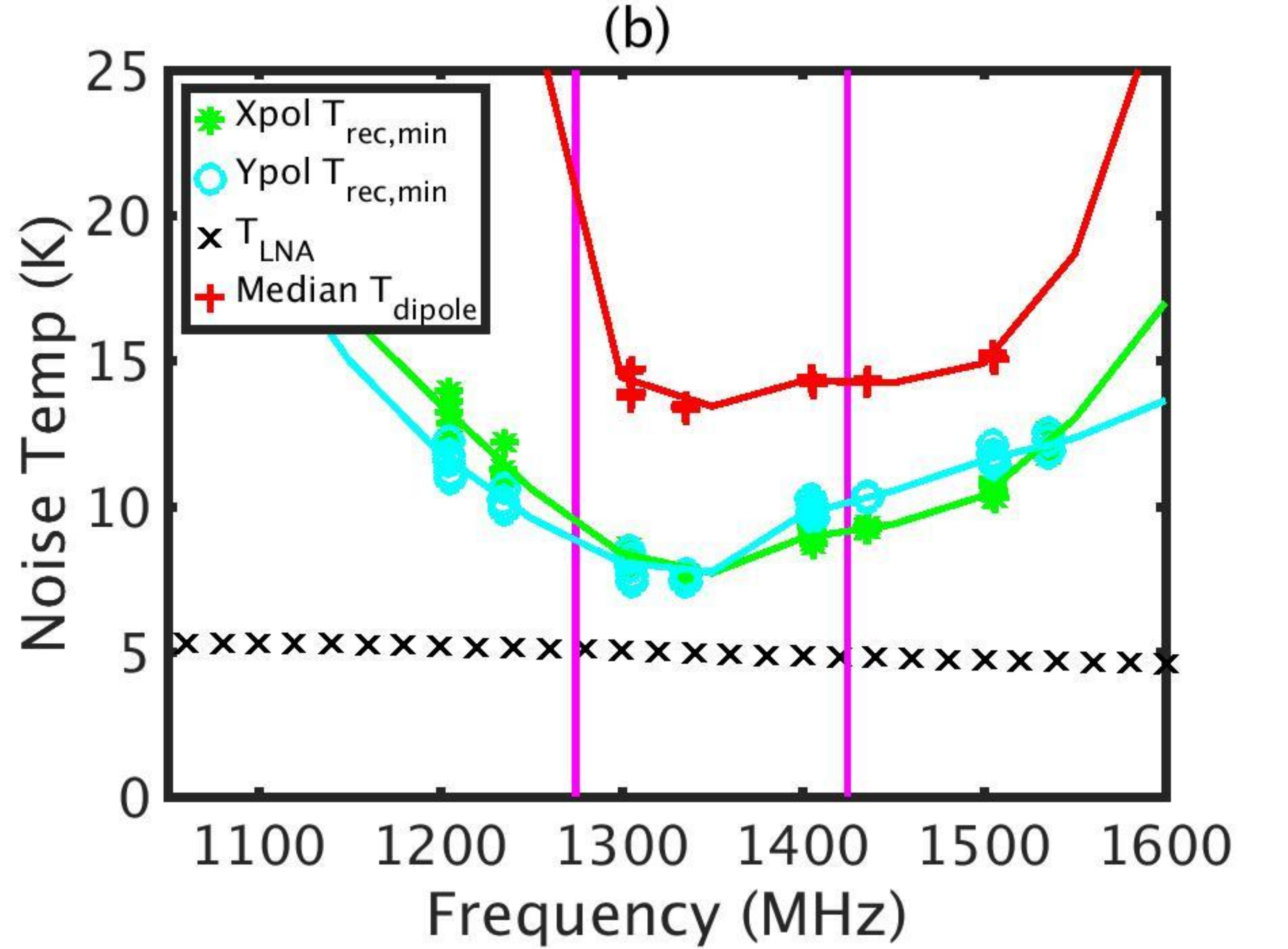}
\caption{(a) Measured receiver temperature for the 37 dipoles near
  1336~MHz.  The dipoles numbered 1 to 19 correspond to
  X-polarization and 20 to 38 correspond to Y-polarization. The LNA
  connected to dipole 13 in Y-polarization was bad.  (b) Median
  receiver temperature versus frequency with an interpolated curve are
  marked in red. The median value is obtained from the 37 dipole
  temperatures.  The measured LNA noise temperature, $T_{LNA}$, is marked with a
  cross. The peak-to-peak spread in the measured LNA temperatures for
  the 37 amplifiers is about 1.5 K. The minimum receiver temperature
  $T_{\rm rec,min}$ (see text) for the X and Y polarizations along with
  interpolated curves are shown in green and cyan. The minimum
  difference between $T_{\rm rec,min}$ and LNA noise temperature is
  about 2.5~K, which we attribute to the losses ahead of the LNA.  The
  vertical lines show the 150~MHz frequency range centered at
  1350~MHz.
\label{fig3}
}
\end{figure}

The receiver temperature of the PAF was measured at the outdoor test
facility at the GBO. This facility is a small outdoor structure with a
retractable roof facilitating hot and cold load measurements with the
PAF receiver to obtain the receiver noise temperature by the Y-factor
method \citep{warnicketal2010}. The receiver is mounted with the
dipoles facing up toward either an ambient temperature load (hot load)
or the cold sky. A large aluminum cone, flaring upward from the edge
of the dipole array, is attached to the receiver and functions to
shield the dipoles from wide-angle ground radiation. To derive the Y
factor, the array output voltage correlation $\bm R_{\rm hot}$ was
measured with an absorber placed in front of the array, which forms an
isotropic hot load at ambient temperature. A second correlation
measurement $\bm R_{\rm cold}$ was made by pointing the array to a
region of sky away from galactic plane, and this formed the cold
load. In this case, the factor
\be
  Y  =  \frac{\bm w^H \bm R_{\rm hot} \bm w }{\bm w^H \bm R_{\rm cold} \bm w},
\label{yeq}
\ee
and the receiver temperature
\be
  T_{\rm rec}  =  \frac{T_{\rm hot} - Y T_{\rm cold}}{Y - 1},
\label{eq1}
\ee
where $T_{\rm hot}$ and $T_{\rm cold}$ are the temperatures of hot
load and cold sky, respectively, and $\bm w$ are the weights. The
derived value of the receiver temperature depends on the weights, as
seen from \eqref{yeq}. The cold sky temperature is estimated from
the sky brightness temperature distribution and typically has a value
of $\sim 8$~K \citep{jean-yvesetal2002}. The hot load temperature is 296.5~K.
Despite the use of the conical ground shield, the cold sky
correlations were not devoid of contribution from ground scattered
radiation. From single feed measurements using the outdoor test
facility, it is estimated that a residual of $\sim 4$~K can be present
in the receiver temperature values due to this scattered radiation.

Fig.~\ref{fig3}a shows the measured receiver temperature $T_{\rm
  dipole}$ for each dipole.  These temperatures are obtained using the
Y-factor derived from the output power of each dipole in the array.
Thus, for example, $T_{\rm dipole}$ for dipole 1 is equivalent to using $\bm w^T = [1,0,0,...,0]$
(i.e. weight 1 for dipole one and 0 for all other dipoles) in
\eqref{yeq} and then estimating the temperature using \eqref{eq1}.
This weighting scheme is applied for each dipole to get 38 $T_{\rm dipole}$
values (see Fig.~\ref{fig3}a). The $T_{\rm
  dipole}$ values range between 12~K and 18~K near 1336~MHz.  The
missing value at dipole 13 in the Y-polarization is due to a bad LNA
on that channel.  The median value of $T_{\rm dipole}$ versus
frequency is shown in Fig.~\ref{fig3}b. The minimum $T_{\rm dipole}$
is about 13.3~K near 1336~MHz.

The measured $T_{\rm dipole}$ has contributions from ground scattering
and mutual coupling between array elements. Since both of these
contributions produce noise correlations at the output of the array,
they can be canceled out to a large extent. This cancellation
corresponds to the maximum Rayleigh quotient of Y, which will provide
a minimum receiver temperature, $T_{\rm rec,min}$. The weight vector
that needs to be applied in \eqref{yeq} to get the maximum Y is the eigenvector
corresponding to the maximum eigenvalue of the matrix $\bm R_{cold}^{-1} \bm R_{hot}$.
Fig.~\ref{fig3}b
shows the $T_{\rm rec,min}$ obtained from X- and Y-polarization data
separately. The measured LNA noise temperature, $T_{LNA}$, versus frequency is
also shown in Fig.~\ref{fig3}b. The minimum difference between the
measured LNA noise temperature and $T_{\rm rec,min}$ is $\sim$ 2.5 K
near 1350 MHz. We attribute this excess noise temperature to
any residual correlated noise and to losses ahead of the LNA, which include (a) loss in the dipoles and
balun, (b) loss in the thermal transition and (c) connection losses.
Thus the measured excess noise temperature
provides an {\em upper limit} to the losses ahead of the LNA (i.e the loss in temperature should
$\le$ 2.5 K). In this paper, we assume that the losses ahead of the LNA
are independent of the beamformer weights.

\section{Observations}
\label{secobs}

\begin{deluxetable*}{ccCr}[b!]
\tablecaption{Observed sources \label{tab1}}
\tablecolumns{4}
\tablewidth{0pt}
\tablehead{
\colhead{Source} & \colhead{RA(2000)} & \colhead{DEC(2000)} & \colhead{S$_{1350 MHz}$} \\
\colhead{} & \colhead{hh:mm:ss} & \colhead{$^{o}\;\;$:$^{'}\;\;$:$^{''}\;\;$} & \colhead{Jy}
}
\startdata
3C123  &  04:37:04.4        &  +29:40:14        &  49.8  \\
3C286  &  13:31:08.3        &  +30:30:33        &  15.4   \\
3C147  &  05:42:36.1        &  +49:51:07        &  22.8 \\
3C295  &  14:11:20.5        &  +52:12:10        &  23.2 \\
3C48   &  01:37:41.3        &  +33:09:35        &  16.7 \\
3C348  &  16:51:08.3        &  +04:59:26        &  49.5 \\
Virgo A &  12:30:49.6        &  +12:23:21        &  218.8   \\
\enddata
\end{deluxetable*}

\begin{deluxetable*}{ccl}[b!]
\tablecaption{Observation log \label{tab2}}
\tablecolumns{3}
\tablewidth{0pt}
\tablehead{
\colhead{Date} & \colhead{Project} & \colhead{Description} \\
\colhead{dd-mm-yy} & \colhead{code} & \colhead{}
}
\startdata
08-03-17 & TEST\_08MAR2017 & Measurement of receiver temperature \\
         &                & LO\tablenotemark{a}=1250,1350,1450,1550,1650,1750 MHz \\
11-03-17 & TGBT17A\_502\_03 & On-Off observations. Sources observed 3C123, \\
         &                & 3C147, 3C48. LO=1350,1450 MHz \\
13-03-17 & TGBT17A\_502\_05 & On-Off observations. Sources observed Virgo A. \\
         &                & LO=1350,1450 MHz \\
13-03-17 & TGBT17A\_502\_06 & Grid observations. Sources observed Virgo A. \\
         &                & LO=1350 MHz \\
14-03-17 & TGBT17A\_502\_07 & On-Off observations. Sources observed 3C123, \\
         &                & 3C147. LO=1250,1350,1450,1550 MHz \\
15-03-17 & TGBT17A\_502\_08 & On-Off observations. Sources observed 3C286, \\
         &                & 3C295,3C348. LO=1250,1350,1450 MHz \\
16-03-17 & TGBT17A\_502\_10 & Sources observed 3C123, B0329$+$54, Rosette Nebula \\
         &                & LO=1350 MHz \\
\enddata
\tablenotetext{a}{Local Oscillator frequency}
\end{deluxetable*}

In March 2017, the phased array receiver was installed on the
GBT. Extensive observations were made on a set of seven calibrator sources
to measure the system performance. A list of observed calibrator
sources, their J2000 coordinates, and flux density at 1350~MHz are
given in Table~\ref{tab1}. A log summarizing observations is given in
Table~\ref{tab2}. The observations can broadly be classified into two
categories: (a) to measure the performance over the FoV; (b) to
measure the boresight performance.  The performance over the FoV was
measured by observing on a grid of positions (`grid' observations)
centered on the strong radio source Virgo A.
The grid observation was made at 1336~MHz (bandwidth $\sim$ 300 kHz).
The grid positions were separated by 3\arcmin~in both the elevation and
cross-elevation directions.  On-source and off-source measurements
were made toward all calibrators in Table~\ref{tab1} to derive the
boresight performance.  The observed off-source positions have
$+1$\ddeg~offset in RA and $0$\ddeg~offset in DEC from the J2000
source positions.  The measurements were made over a set of
frequencies ranging from 1200 MHz to 1500 MHz, each with a bandwidth
of 300~kHz.

\section{Results}
\label{result}

\begin{figure}[!t]
\plotone{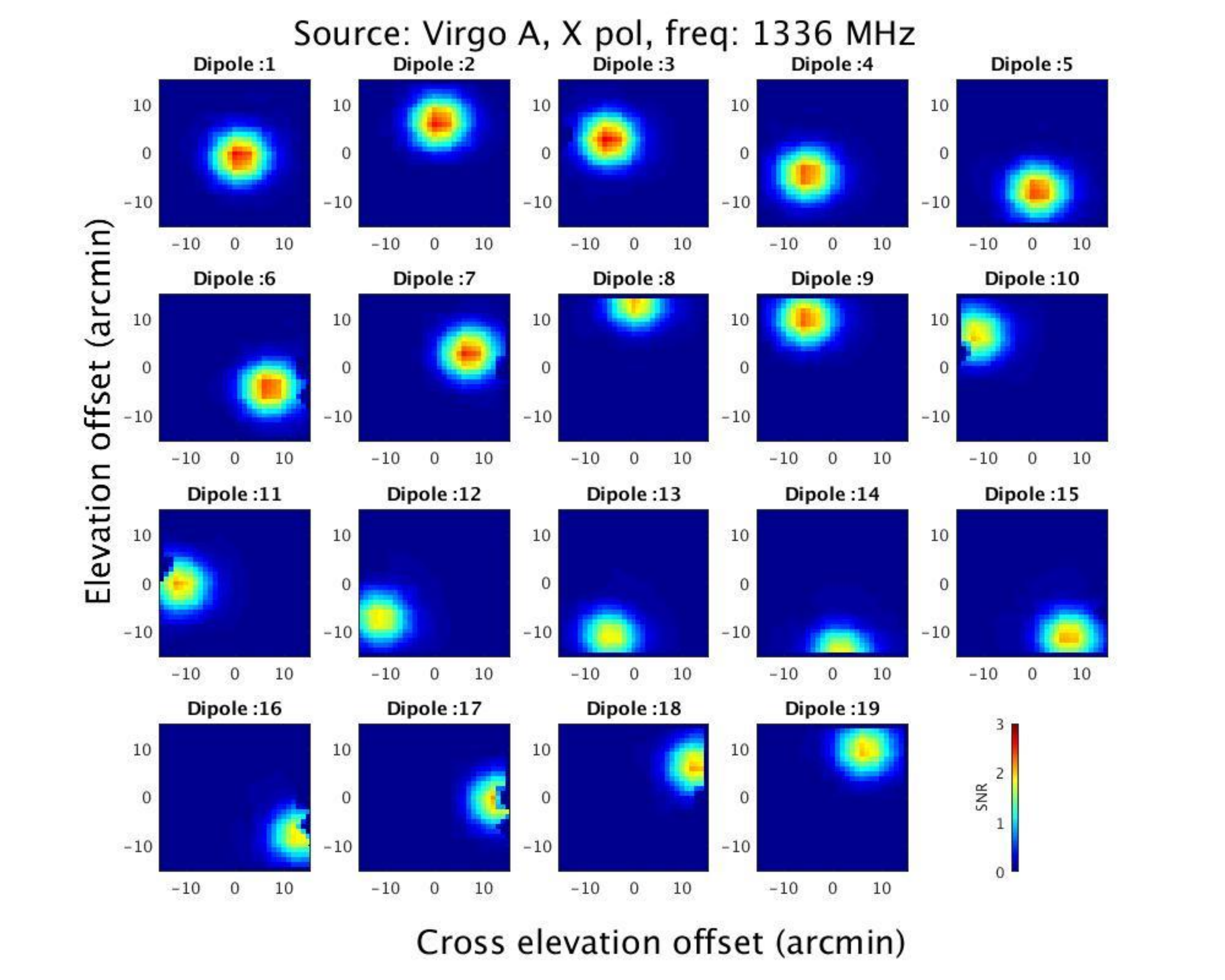}
\caption{
The S/N distributions at 1336 MHz in elevation and cross-elevation
directions for each X polarization dipole obtained from grid observations.
}
\label{fig5}
\end{figure}

\begin{figure}[!t]
\plotone{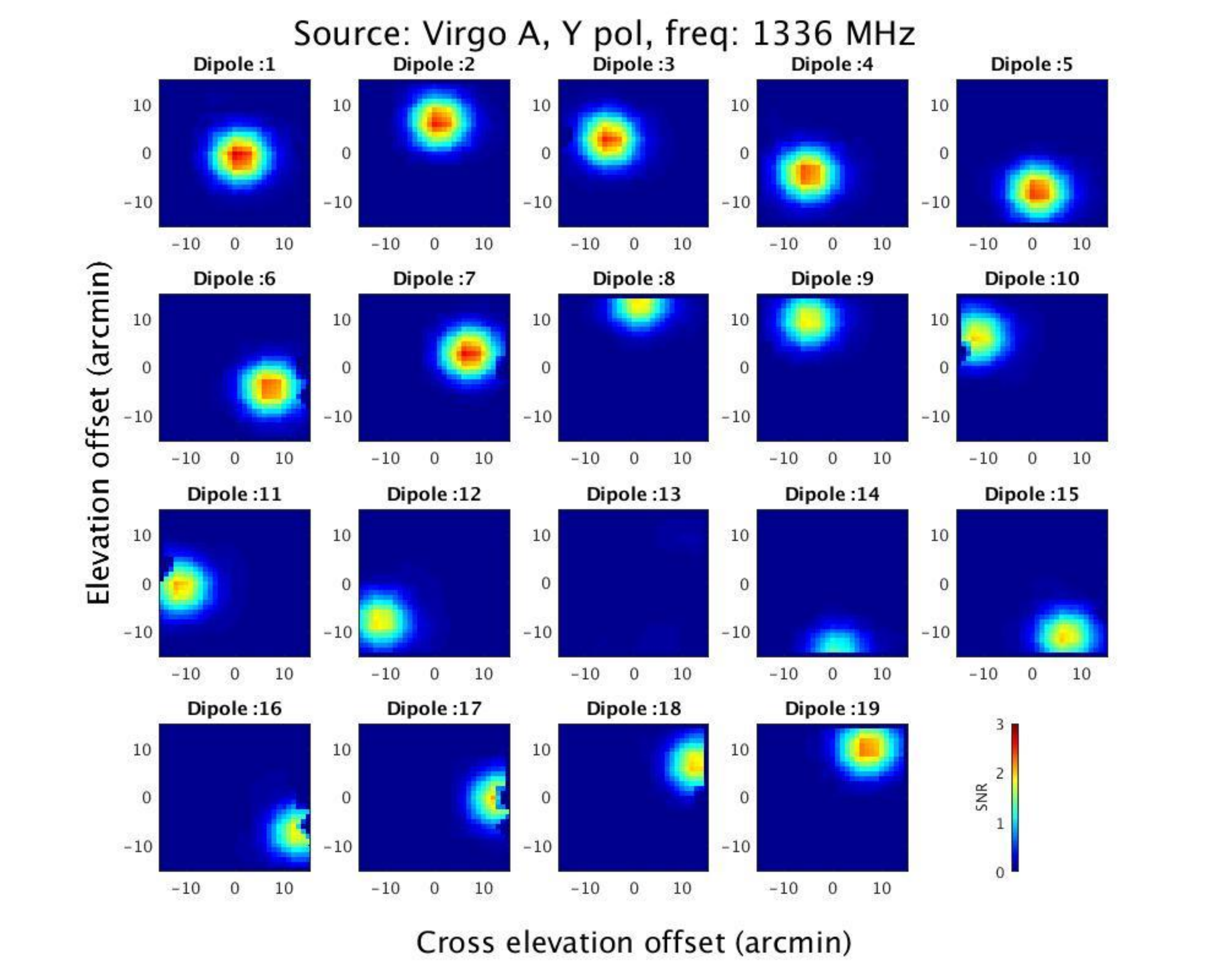}
\caption{
Same as Fig.~\ref{fig5} but for Y polarization. Dipole 13 has a faulty
LNA and dipole 14 has peak S/N a factor of 2.0 lower than that of the
central dipole due to a faulty digital link.
}
\label{fig6}
\end{figure}

The grid observations are used to check the response of individual
dipoles.  Figs.~\ref{fig5} and \ref{fig6} show the distribution of the
S/N (as defined in \eqref{snr}) but for a single dipole) obtained
on Virgo A from each of the 19 $\times$ 2 dipoles (i.e. without
forming beams).  The peak S/N for the central dipole was about 2.6. As
noted earlier, the dipole number 13 in Y-polarization was not
functional. Dipole 14 in Y-polarization had a lower peak S/N compared
to the central dipole by a factor of 2.  This lower S/N is attributed
to a faulty digital link. As discussed below, the two faulty dipoles
have affected the results obtained from the Y-polarization data set.

\subsection{$T_{\rm sys}/\eta$ over the FoV}

\begin{figure}[!t]
\plottwo{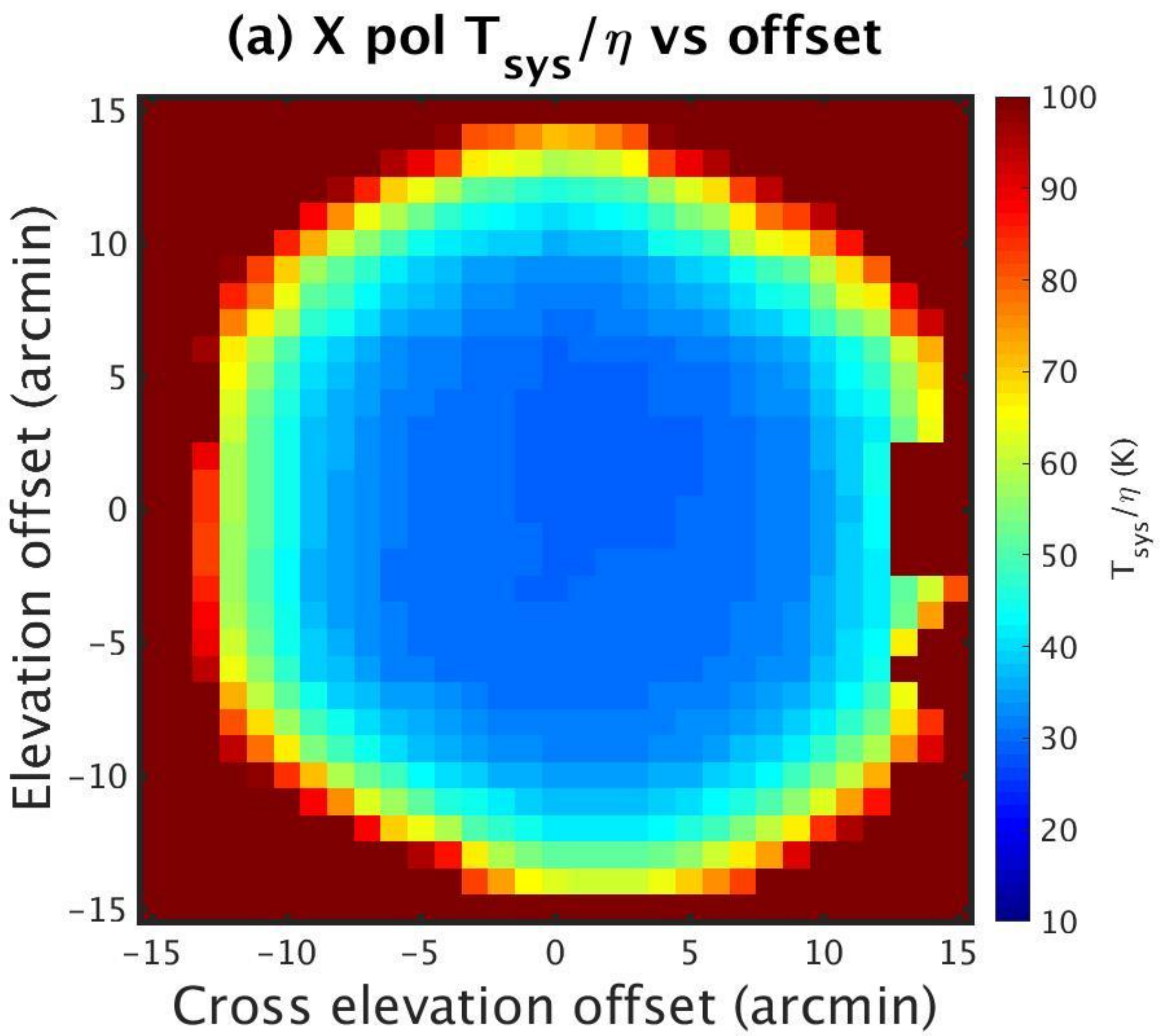}{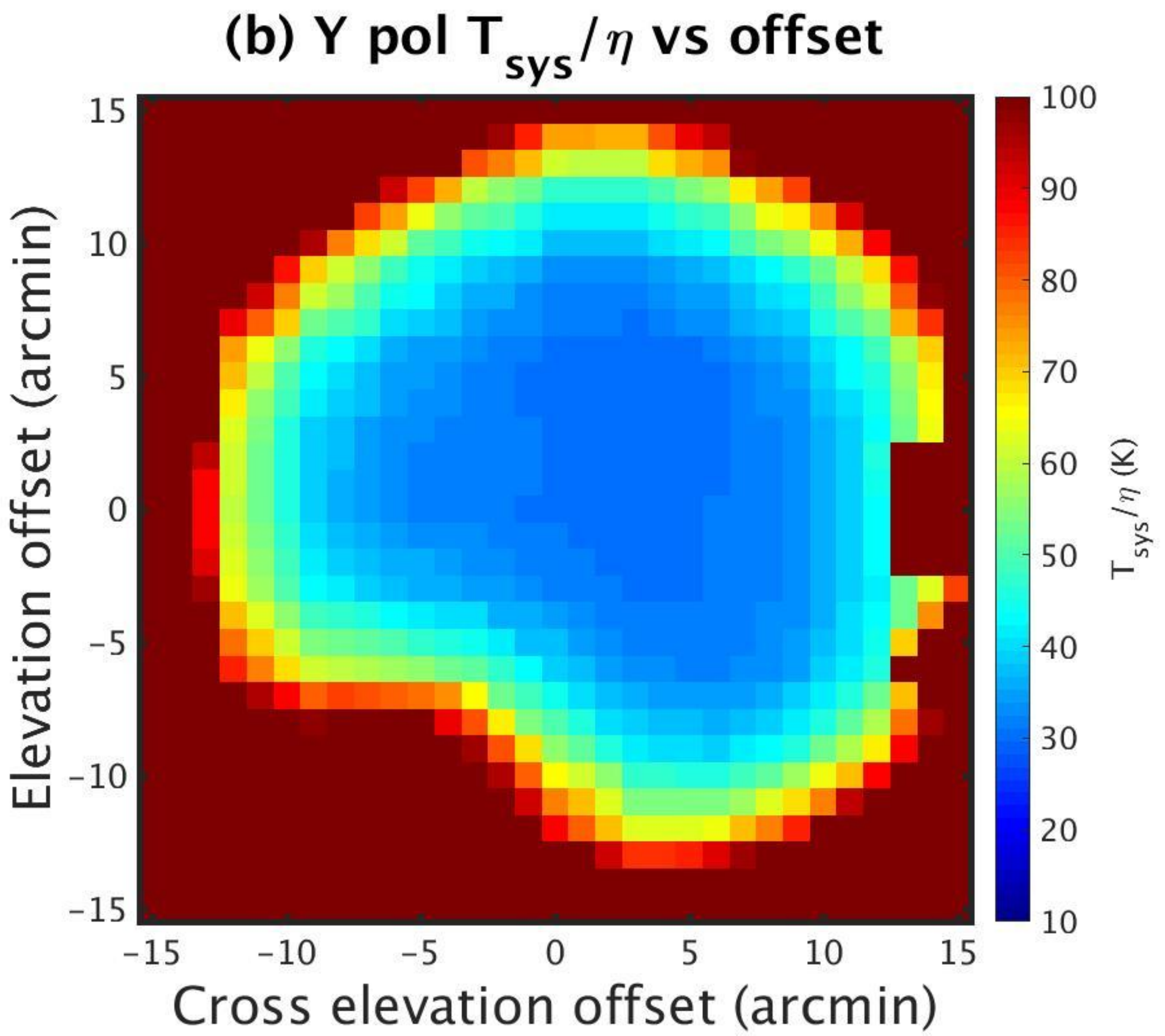}
\caption{(a) Distribution of the $T_{\rm sys}/\eta$ in elevation and
  cross elevation directions for X-polarization obtained from Virgo A
  grid observations.  (b) Same as (a) but for Y-polarization. The
  asymmetry seen in the south-west is due to faulty dipoles 13 and 14
  (see text).
\label{fig7}
}
\end{figure}

\begin{figure}[!t]
\plottwo{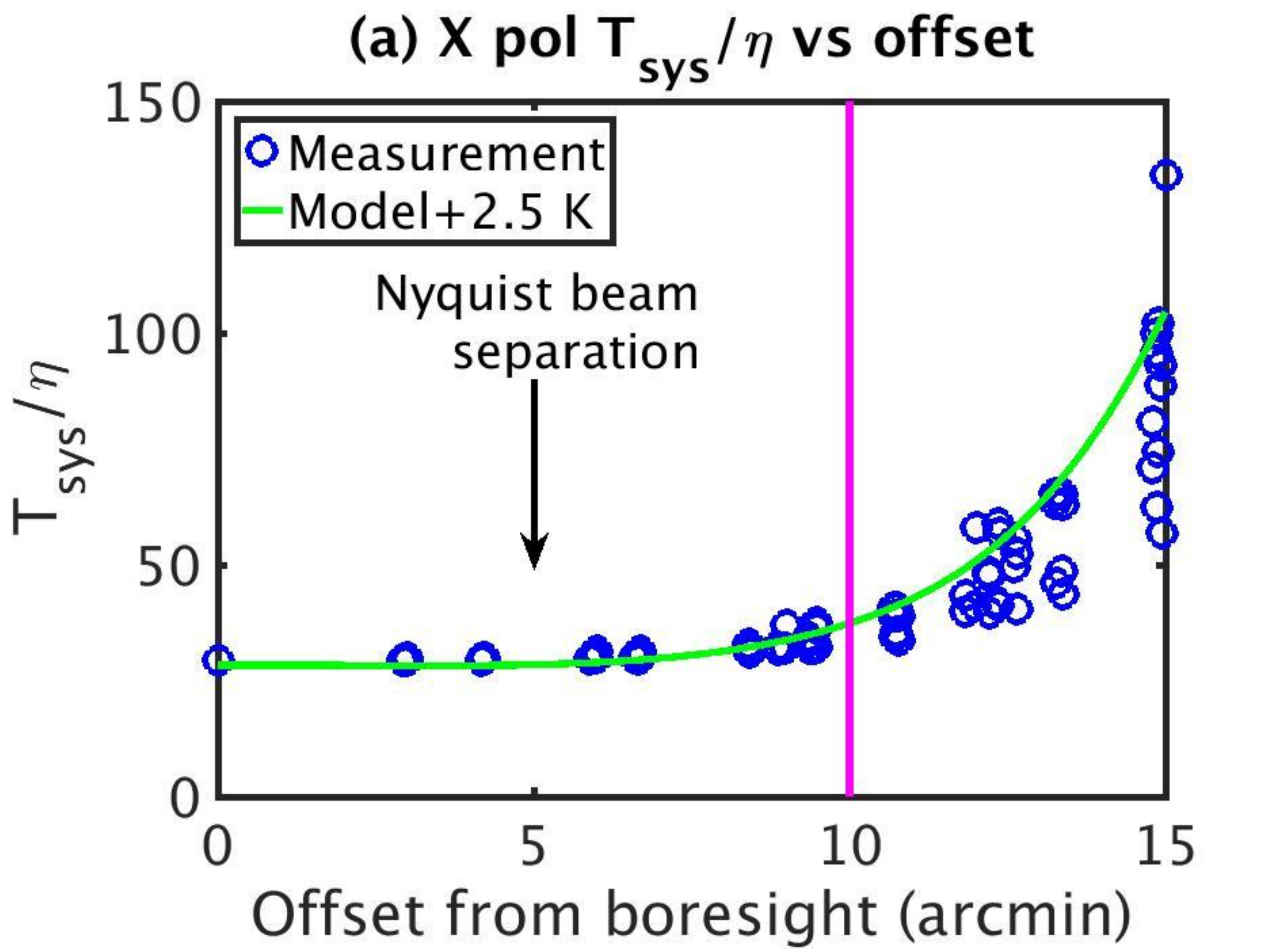}{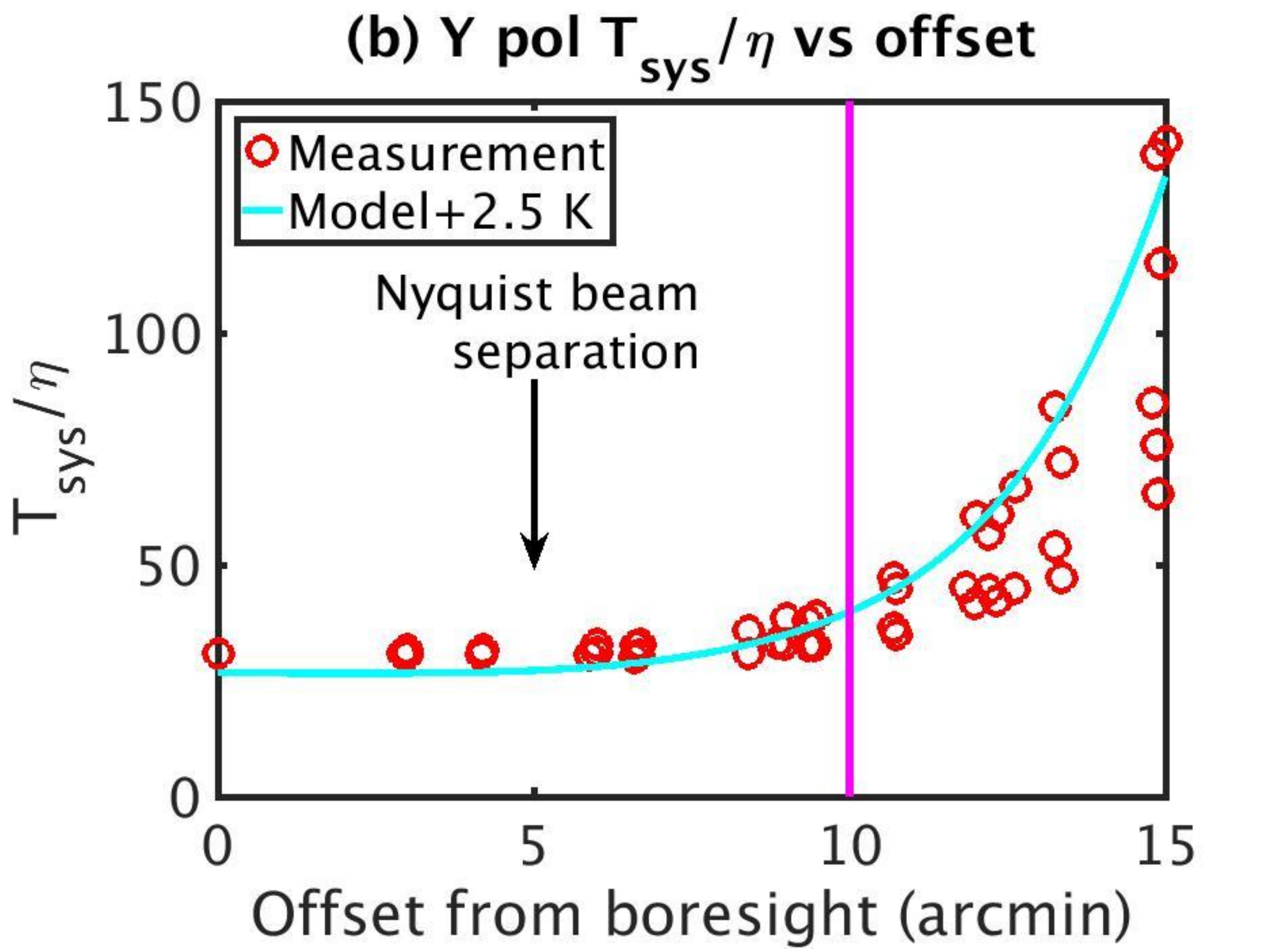}
\caption{(a) The $T_{\rm sys}/\eta$ versus radial offset from the
  boresight for X-polarization are marked in circles. The half power
  beam width of 10\arcmin\ at 1336~MHz is marked by the vertical
  line. The PAF model prediction is shown by the solid line.  The
  system temperature in the model is increased by 2.5~K to account
  for the loss ahead of the LNA (see Fig.~\ref{fig3}b). (b) Same as (a) but for
  Y-polarization. Data points in Fig.~\ref{fig7}b with elevation
  offset $>-3$\arcmin~are used for making the Y polarization plot.
\label{fig8}
}
\end{figure}

\begin{figure}[!t]
\plottwo{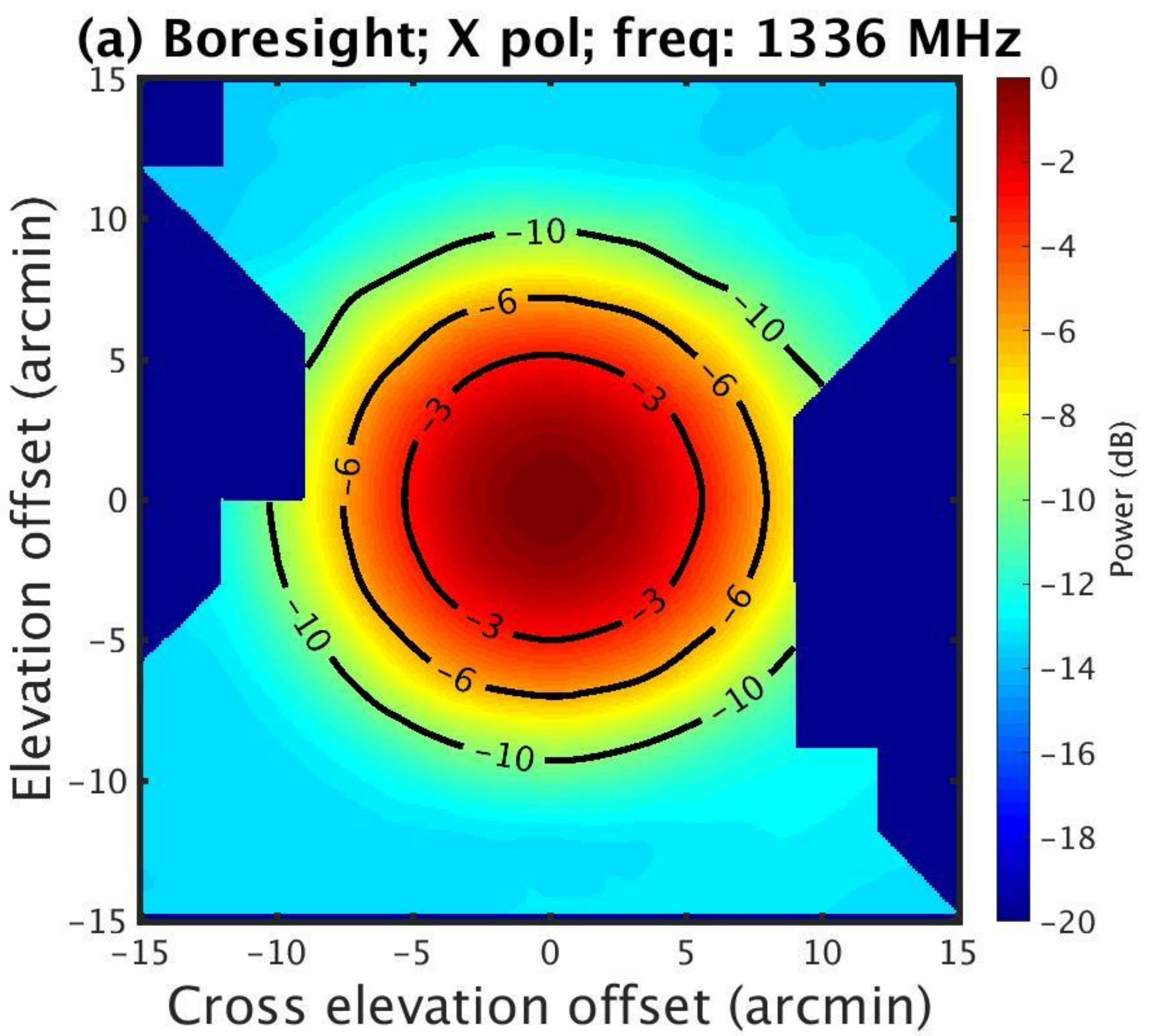}{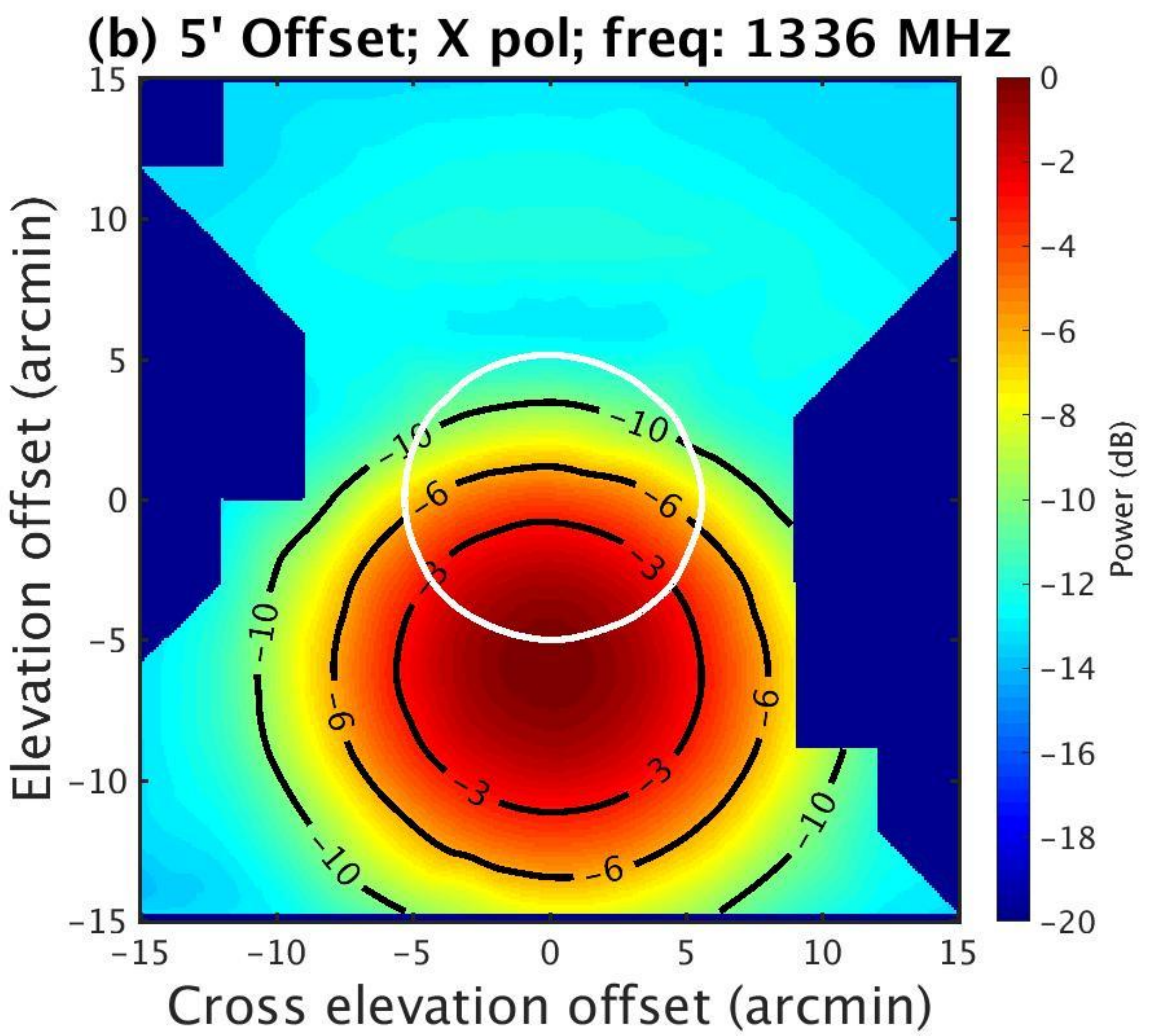}
\caption{
(a) Boresight beam formed using maximum S/N weights. The -3,-6 and -10
  dB contours are marked. (b) An offset beam at about 5\arcmin\
  south of the boresight direction. The -3,-6 and -10~dB
  contours are marked in black. The contour in white is same as the -3~dB
  contour of the boresight beam shown in (a). The dark blue regions on the
  left and right side of the plots are due to loss of data resulted from
  the applied cubic interpolation to make the figures. }
\label{fig9}
\end{figure}

A map of the $T_{\rm sys}/\eta$ is obtained from the grid observation
data set by maximizing the S/N at each offset
position. Figs.~\ref{fig7}a and \ref{fig7}b show the distribution of
$T_{\rm sys}/\eta$ as a function of elevation and cross-elevation
offsets. The distribution is fairly symmetric about the center for the
X polarization. The asymmetry seen in the Y polarization towards the
south-west side is due to the faulty dipoles 13 and 14.  The radial
distributions (i.e. offset from the boresight) of the normalized
$T_{\rm sys}/\eta$ for X and Y polarizations are shown in
Figs.~\ref{fig8}a and \ref{fig8}b.  Only data points for elevation
\gsim~$-3$\arcmin~from those shown in Fig.~\ref{fig8}b are used for
making the radial distribution of Y polarization.  The half power
beam width at 1336 MHz is $\sim$ 10\arcmin. The beam separation
required for Nyquist sampling (i.e. $\lambda/(2D)$, where
$\lambda$ is the wavelength of observation and $D=100$~m is the
diameter of the GBT aperture plane; \cite{padman1995})
is $\sim$ 4\arcmin. The
degradation in $T_{\rm sys}/\eta$ at this offset is $\sim 1$\%.

The grid observations were used to examine the seven formed beam patterns.
Fig.~\ref{fig9}a shows the boresight maximum S/N beam measured using
Virgo A and in Fig.~\ref{fig9}b a beam with 5\arcmin~offset from
boresight is shown.  The maximum S/N beams are smooth and
approximately Gaussian for level between 0 and -10 dB. All of the seven
formed beams show similar properties.

\subsection{Boresight $T_{\rm sys}/\eta$}
\label{boresight}

\begin{figure}[!t]
\plottwo{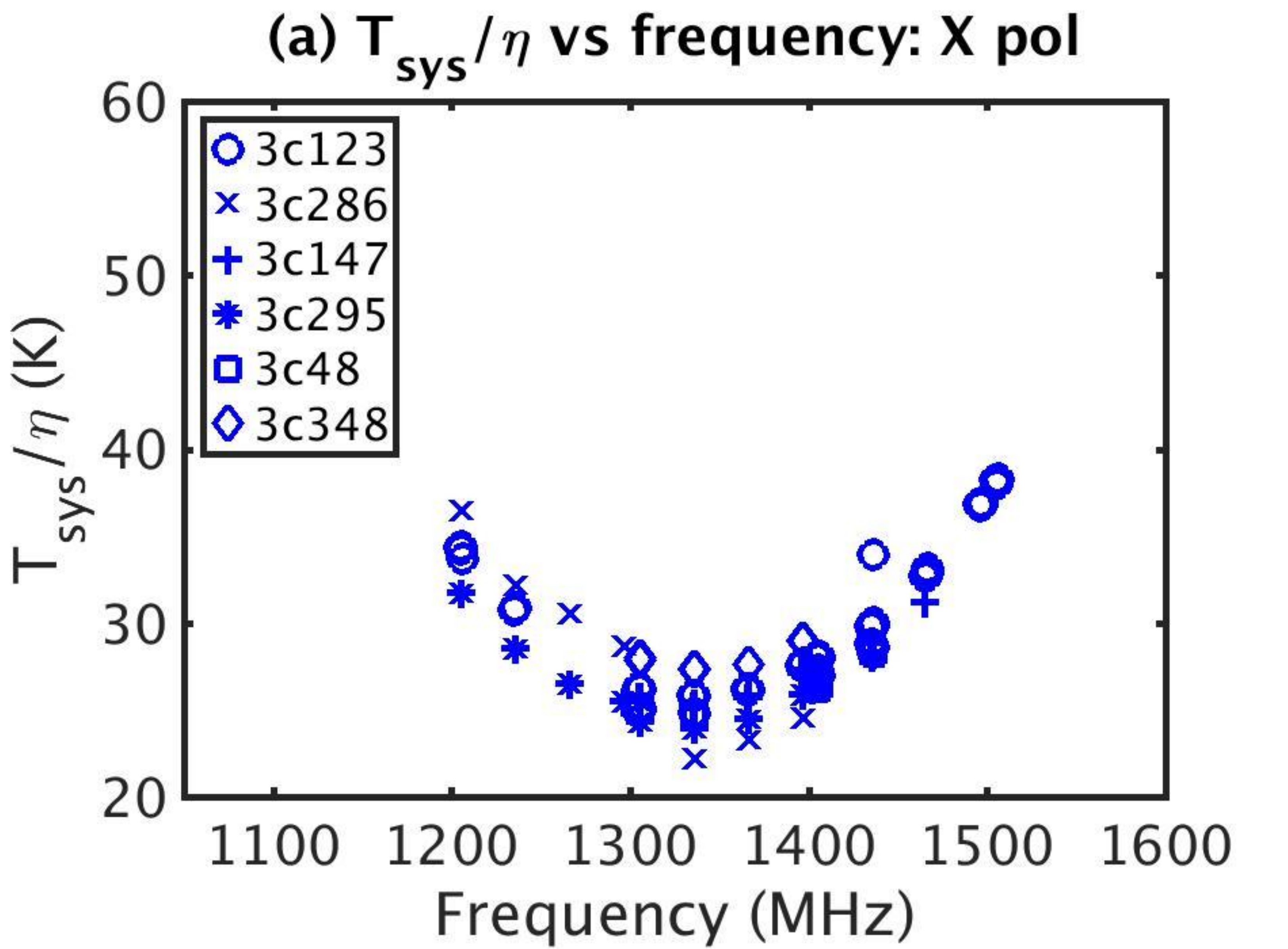}{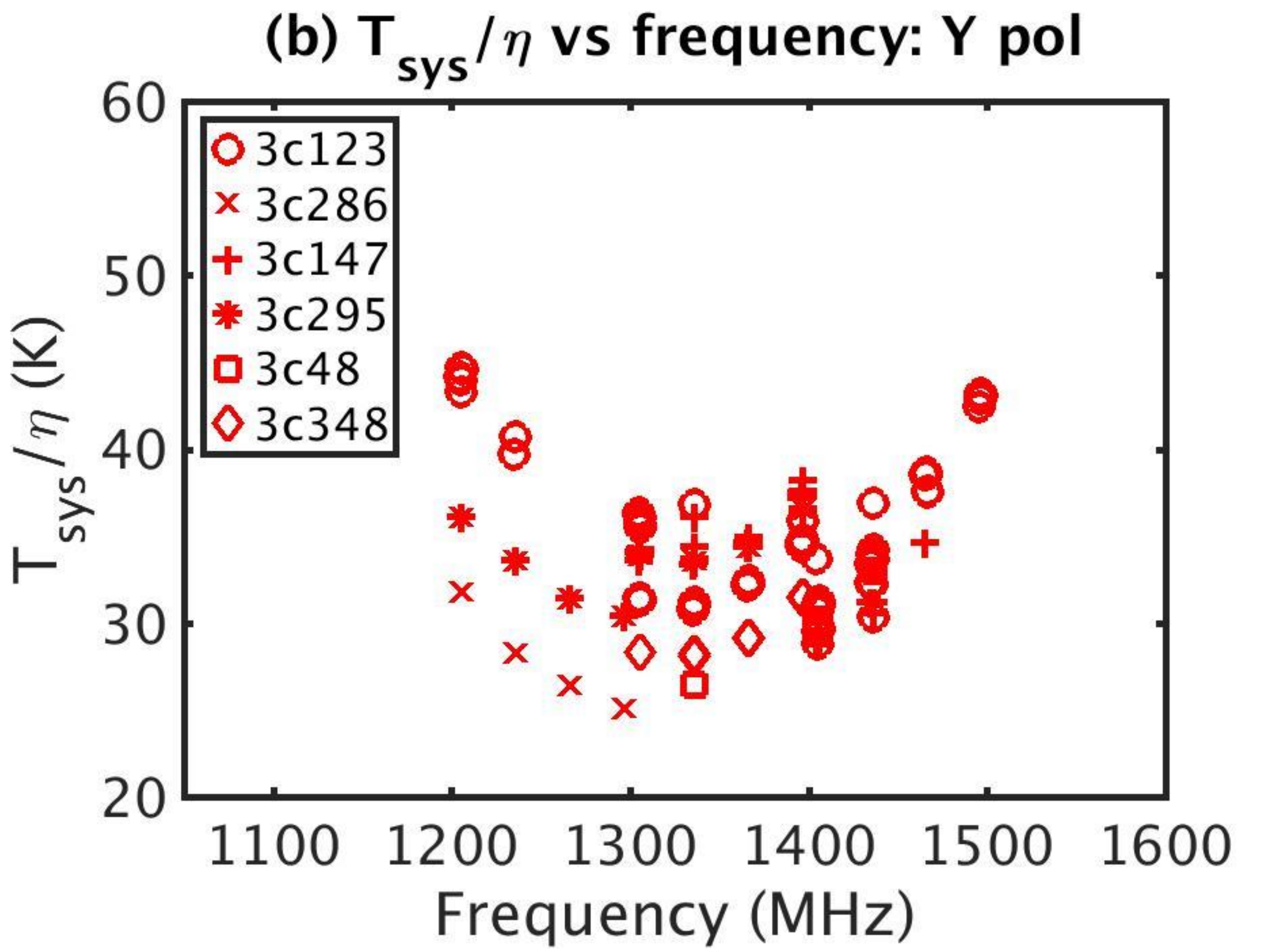}
\caption{The $T_{\rm sys}/\eta$ for boresight beam obtained from
  observations with the PAF on the GBT for
  X-polarization (a) and  Y-polarization (b). The sources observed are
  indicated in the legend.
\label{fig10}
}
\end{figure}

\begin{figure}[!t]
\plottwo{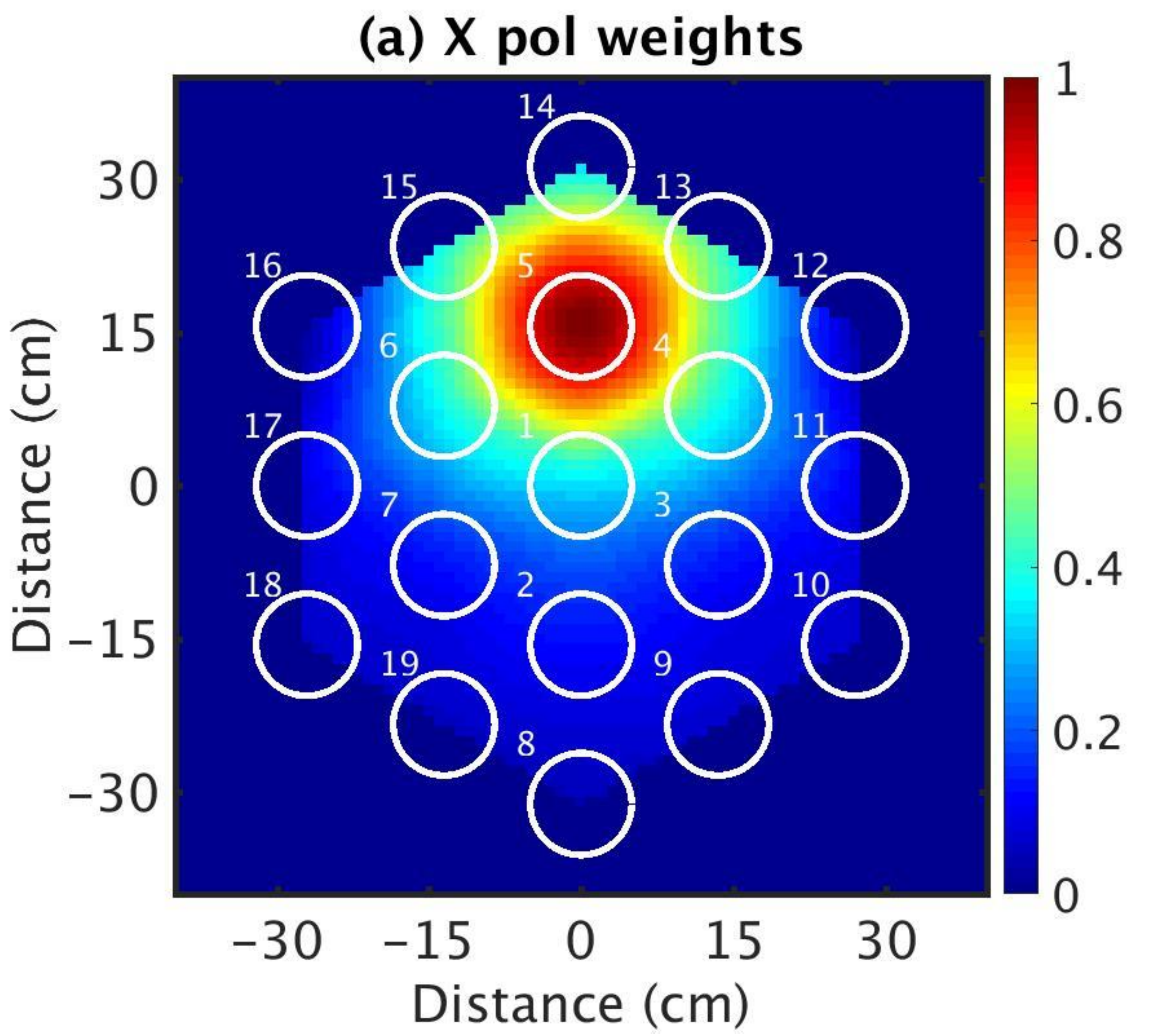}{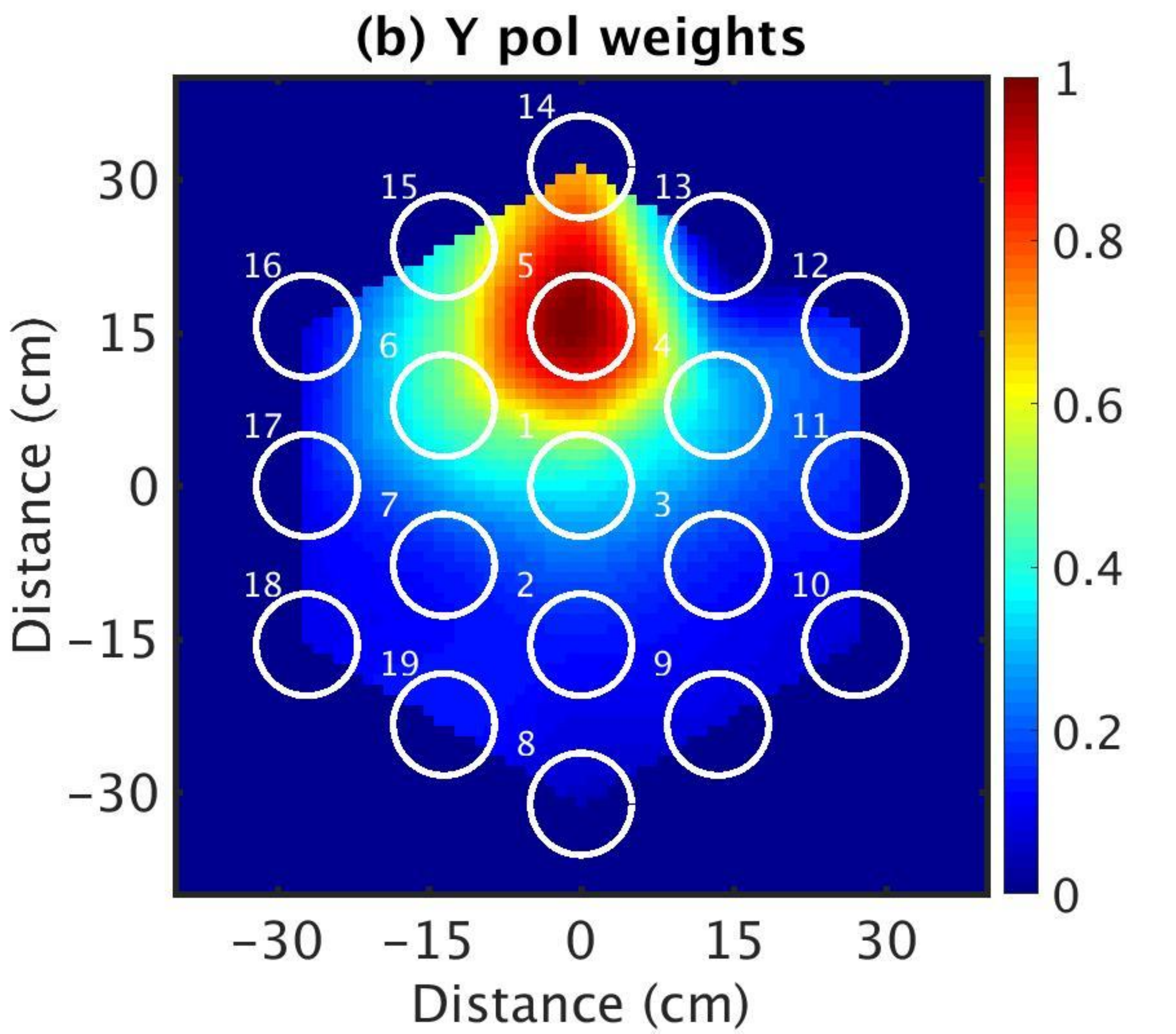}%
\caption{ The normalized weight distributions obtained from 3C147
  observations near 1336~MHz for X and Y polarizations are shown in
  (a) and (b). The distributions are superposed on the dipole array
  geometry; the white circles show the location of the dipoles.
  These plots illustrate the reason for higher $T_{\rm sys}/\eta$ and larger
  spread in the measured values for Y polarization. The $T_{\rm sys}/\eta$ for the Y-polarization obtained
  from the 3C147 data set is 36~K and that for X-polarization is 25~K (see
  Fig.~\ref{fig10}).  The distribution of weights is centered on
  dipole 5 due to the telescope pointing offset. Dipoles 13 and 14
  (both are faulty dipoles) in Y-polarization are needed to get the
  optimum $T_{\rm sys}/\eta$. Thus the higher $T_{\rm sys}/\eta$ for the Y
  polarization compared to the X-polarization in this measurement is
  due to the faulty dipoles.
\label{fig11}
}
\end{figure}

\begin{figure}[!t]
\plottwo{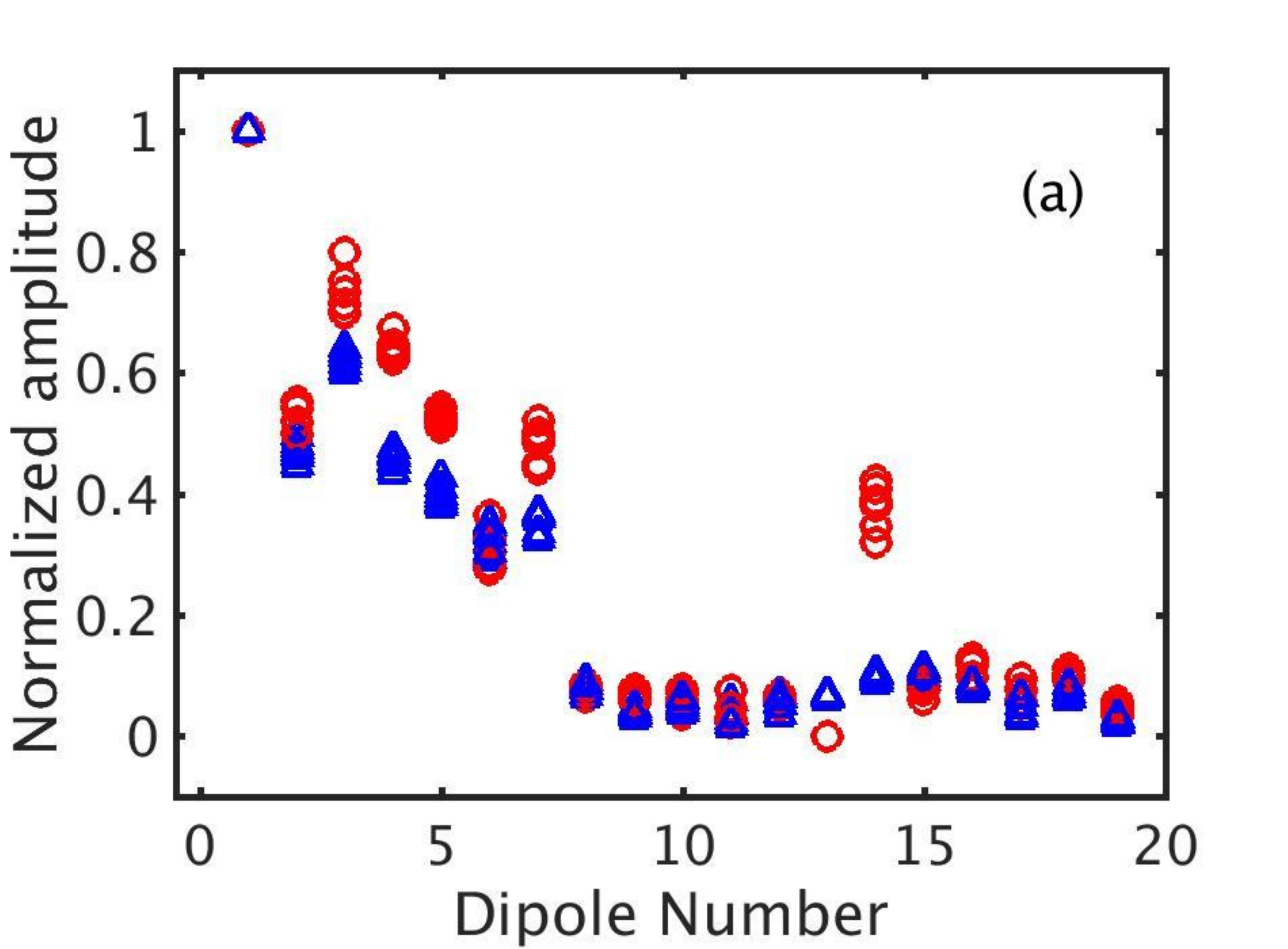}{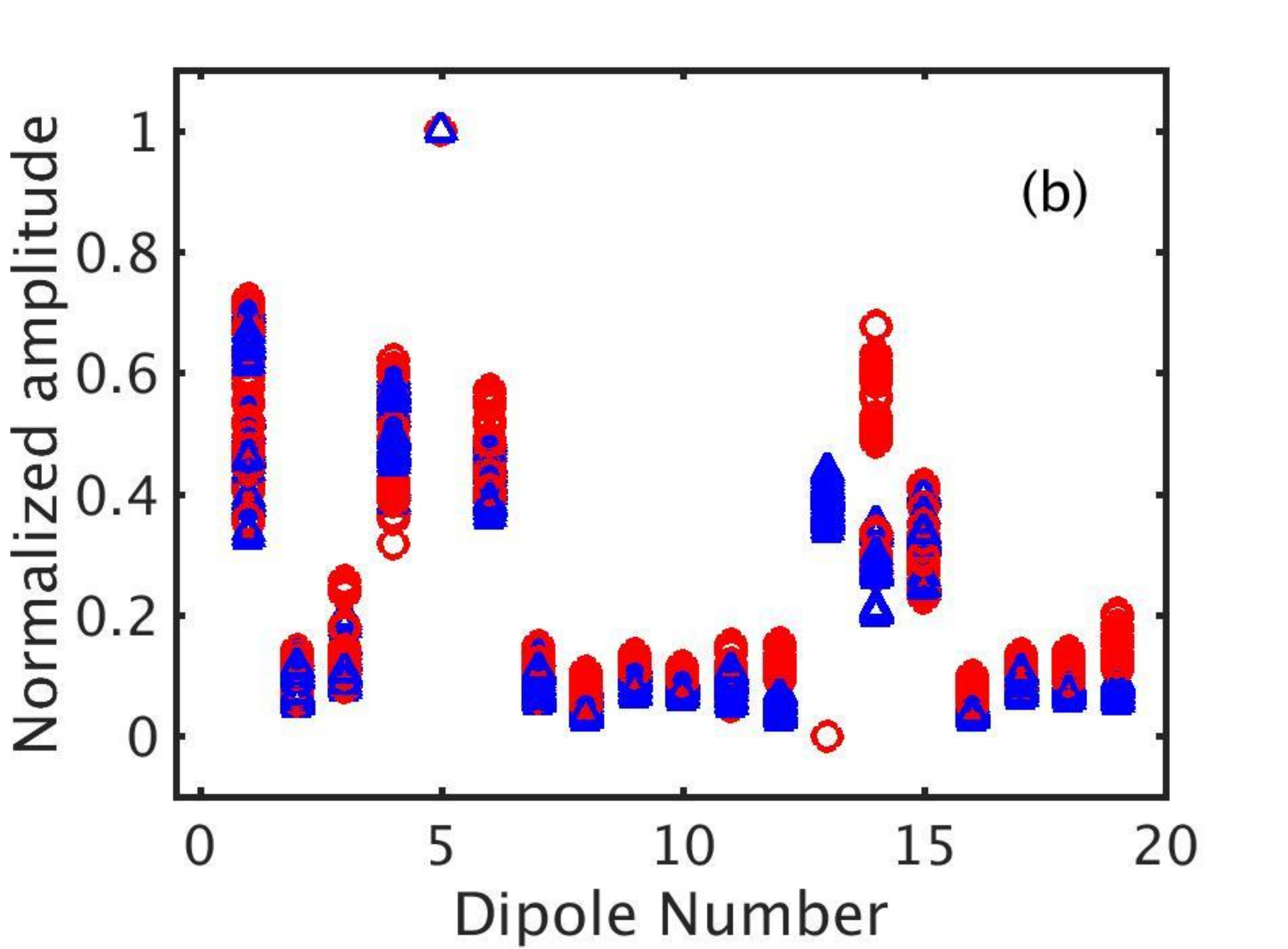}%
\caption{
(a) Normalized weights (blue - X polarization; red - Y polarization) obtained from a subset of
  measurements shown in Fig.~\ref{fig10} where the weight distribution
  is centered at dipole 1. For this subset of measurements the $T_{\rm
    sys}/\eta$ for both X and Y polarizations are comparable. (b)
  Normalized weights for the subset of data with $T_{\rm
    sys}/\eta > 30$ K for Y-polarization in the frequency interval
  1300 to 1400 MHz (see Fig.~\ref{fig10}b). The weight distribution
  for all these measurements are centered near dipole 5.
\label{fig12}
}
\end{figure}

Plots of the measured boresight $T_{\rm sys}/\eta$ versus frequency for
the two polarizations are shown in Figs.~\ref{fig10}a and
\ref{fig10}b.  The data presented in these plots are from observations
toward sources with flux density $<50$~Jy. The weights for
beamforming were derived by maximizing S/N on the observed source
itself. The best median $T_{\rm sys}/\eta$ for X polarization is 25.4~K at
1336~MHz and the peak-to-peak spread in the values is $\pm 2.5$~K.
The scatter in X-polarization values are due to a combination of:
(a) uncertainty in the flux densities of the sources; (b) variation in
off-source sky contribution; (c) telescope pointing offset. The
3$\sigma$ thermal noise uncertainty in all these measurements is
$\sim 0.3$~K. The median $T_{\rm sys}/\eta$ increases by $\sim 5$~K near
the edge of the 150~MHz bandwidth of interest, centered at 1350~MHz.

The median $T_{\rm sys}/\eta$ of Y polarization near 1336~MHz is
$32.3 \pm 5$~K (peak-to-peak).  To investigate the origin of the higher
$T_{\rm sys}/\eta$ and the larger scatter of the Y-polarization
measurement, we examine the observation toward 3C147 near 1336~MHz.
The $T_{\rm sys}/\eta$ derived from this data set are 25 and 36~K
for the X- and Y-polarizations, respectively.  The normalized weight
distributions obtained from this data set are shown in
Fig.~\ref{fig11}a and~\ref{fig11}b. As seen in the figure, the weight
distributions are centered near dipole 5, which is due to an
uncorrected telescope pointing offset. Since dipole 13 and 14 in
Y-polarization are faulty, the PAF is unable to form an optimum beam
and hence the $T_{\rm sys}/\eta$ for the Y polarization is higher. We
examined the data sets from all calibrators in the frequency range
1300 to 1400~MHz and found that a subset of the observations have
telescope pointing offsets close to zero as inferred from the weight
distribution (see Fig.~\ref{fig12}a). The derived $T_{\rm sys}/\eta$
for this subset are similar for both polarizations and have values
between 25 and 28~K. The weight distribution for the subset with
$T_{\rm sys}/\eta$ $>30$~K for Y-polarization is shown in
Fig.~\ref{fig12}b. As seen in the figure, the weight distribution is
centered at dipole 5 for all the data in this subset. We conclude that
the larger scatter and higher $T_{\rm sys}/\eta$ for Y-polarization
compared to X-polarization is due to the telescope pointing offset and
the presence of the two faulty dipoles. The large telescope pointing
offset is because the GBT did not have an accurate pointing model for
the PAF receiver system while we were doing the measurements.

The radio source Virgo A is the strongest calibrator source observed
during the commissioning. The flux density of this source is 218.8~Jy
at 1350~MHz. The best $T_{\rm sys}/\eta$ measured on this source are
28 and 30 K for the X and Y polarizations respectively. The 1$\sigma$
noise is 0.3~K. These values are about 3 to 5~K higher than the median
value obtained from sources with flux density $<50$~Jy. A possible
cause of this higher $T_{\rm sys}/\eta$ is the noise contribution due
to the source itself \citep{ananthaetal1991}, which may be affecting
the beamformer weight solutions.

We compare the performance of the PAF with the existing, cryogenic,
optimized single feed 1.4~GHz receiver on the GBT, which has a
$T_{\rm sys}/\eta$ $\sim$
25.7~K\footnote{\url{https://science.nrao.edu/facilities/gbt/proposing/GBTpg.pdf}}.
Our measurements show that the performance of the PAF system is
comparable to this single feed receiver. Achieving comparable
performance is a major milestone in the development of the
PAF. Table~\ref{tabss1} lists the $T_{\rm sys}/\eta$ of multi-beam
receivers in other telescopes for comparison.

\subsection{Survey speed}

{


The intrinsic survey capability per unit bandwidth of an astronomical receiver is found by evaluating the squared sensitivity integral given in \citep[Eq.~83]{haybird2015},
\begin{equation}\label{eq:ssfom_hay}
  \mbox{SSFoM} = \int \mathcal{S}^2(\Omega) \, d\Omega
\end{equation}
where the sensitivity map  $\mathcal{S}(\Omega)$ is the receiver sensitivity as a function of sky angle $\Omega$.  By dividing the survey speed figure of merit, SSFoM, by the peak sensitivity, the survey speed weighted field of view, SSFoV, of the instrument can be found as
\begin{equation}
  \mbox{SSFoV} = \frac{1}{\mathcal{S}^2_\text{max}} \int \mathcal{S}^2(\Omega) \, d\Omega
\label{eqssfov}
\end{equation}
The PAF sensitivity map is measured by steering the telescope to place a bright calibrator source at each of a grid of closely spaced points in the sky, measuring the realized sensitivity of the receiver with the source at that location, and thereby sampling the sensitivity map at many discrete points. The integrals in SSFoM and SSFoV are approximated as sums over the sample points. These figures of merit are quite general, and can be used to compare single-pixel receivers, cluster feeds, phased array feeds, and aperture arrays on an equal footing.

To remove the dependence of the field of view on dish size, and obtain a dimensionless number that allows instruments with different aperture sizes to be compared more easily, the survey speed weighted field of view is commonly expressed as a number of beams. The angular field of view is divided by the area of the sky per beam when fully sampled by independent, formed beams to obtain 
\begin{equation}\label{eq:nb}
  N_b = \frac{\mbox{SSFoV}}{\Omega_b} = \frac{\mbox{SSFoM}}{\Omega_b\; \mathcal{S}_\text{max}^2}
\end{equation}
The amount of beam overlap required for full sampling, which determines the value of the area $\Omega_b$ per fully sampled beam, is treated in \cite{haybird2015}. 
While \eqref{eq:ssfom_hay} is the primary definition of SSFoM, the number of beams in \eqref{eq:nb} is quite convenient and widely used in comparing various types of instruments.

\subsubsection{Practical survey speed with real time beamforming}
\label{ncorr}

The field of view expressed as a number of independent beams $N_b$ as in \eqref{eq:nb} is generally not equal to the number of beams that are formed in signal processing. The above considerations apply to the intrinsic survey speed of the analog receiver front end. When the digital back end is included in the analysis, the survey speed of an instrument may be limited further in the signal processing when the receiver back end operates in real time beamforming mode. To reduce hardware costs, fewer beams than are required to fully sample the field of view may be formed. In this case, the practical survey speed of the instrument is lower than the intrinsic SSFoM in \eqref{eq:ssfom_hay}.

In other cases, such as array receivers with signal processing back ends that allow post-correlation beamforming, in certain observing modes more beams may be formed than are required to fully sample the field of view. In this case, the number of formed beams is greater than $N_b$ in \eqref{eq:nb}. If the formed beams overlap, there is correlation between the signal and noise from beam to beam and the information provided by the beams may not be independent.

In view of the complex relationship between the field of view expressed as a number of beams and the number of beams that are formed in digital signal processing, there are differences in the assumptions made in the reported number of beams from one instrument to another. Despite the ambiguity, field of view as number of beams is so convenient a metric, and so generally used in the community, that we accept the disadvantages of this way of parametrizing field of view and provide comparison values, while stating as carefully as possible the assumptions made in the calculation.

To analyze the practical survey speed with a given number of formed beams, we consider three observing cases, which are typical applications of a PAF system on a telescope.
\begin{enumerate}
\item
Survey observations that do not require averaging of data simultaneously
obtained from the different beams.
Examples for this observing case are searching for pulsars and \HI\ observations
of objects with angular size smaller than the FWHM beam size.

\item
Imaging a region of the sky with the PAF system. \HI\ and continuum imaging of extended
sources are examples for the second case. Performing such observations with the PAF in
the On-The-Fly (OTF) mode has some advantages, for example, to reduce the telescope
pointing overhead. Averaging data obtained from different beams in
an OTF observation can optimize the survey speed.
However, the sky position needs to be aligned before averaging the
images from the different beams, which makes the noise in each pixel
of the images un-correlated.  This is demonstrated in
Fig.~\ref{fig13xy} using a 1D scan toward Virgo A obtained with FLAG.

\item
Some observations require averaging of data simultaneously
obtained from the different beams.
An example for such case is imaging of \HI\ emission from an extended source and then
smoothing the image to obtain a low angular resolution ($>$ FWHM beam width) spectrum.
The noise correlation between beams will result in a
spatial correlation of noise in the image and hence the S/N will not improve
by the square root of the number of pixels averaged. As discussed below, 
this class of observations will benefit by keeping the PAF
beams at twice the Nyquist separation or more. 

\end{enumerate}
Survey speed metrics for each of these observation cases will be given in the next section.

\subsubsection{Effective number of beams}

In a PAF system, if a large number of highly overlapping beams were formed in signal processing, 
the beams would be highly correlated, and little new information could be gleaned from adjacent beams. This manifests as strong signal and noise correlation between beams. Further, the sensitivity of each formed beam will differ because of the correlation of receiver noise between elements in the PAF. Following closely the intrinsic number of beams defined by \eqref{eq:nb}, we develop in this section a figure of merit, in a practical sense, that measures the field of view of the PAF with a given number of formed beams, the {\em effective number of beams}.

\begin{figure}[!t]
\plottwo{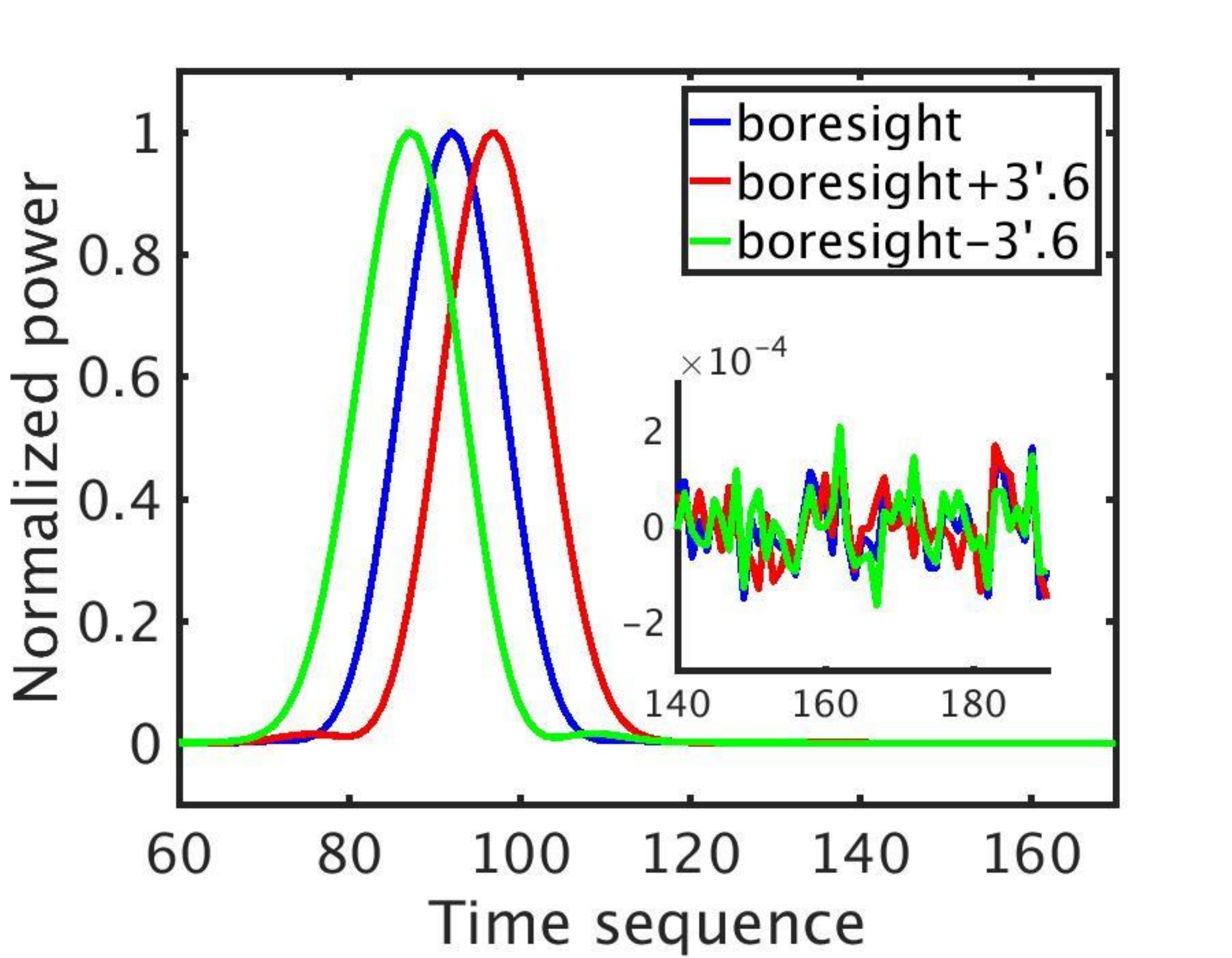}{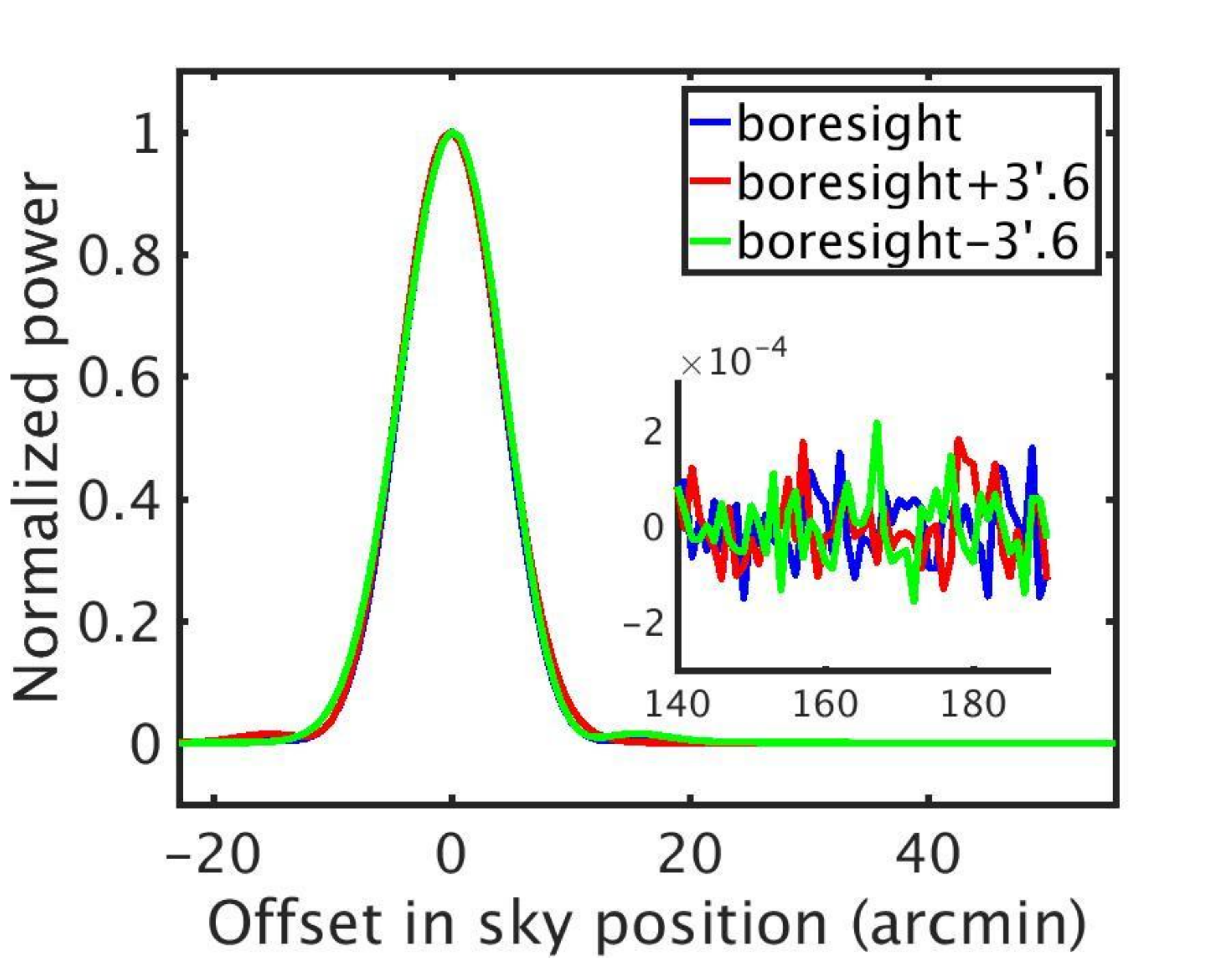}%
\caption{
(a) Normalized power vs time from the three beams of the PAF in an observing scan
made towards Virgo A. The two
off-boresight beams are separated by about 3\arcmin.6 (approximately Nyquist spacing)
on either side of the boresight beam. The inset shows the data from the three beams in the
off-source region. The correlation of the noise fluctuations from the three
beams are evident. The axes of the inset have the same units as the main figure axes.
(b) Normalized power vs sky position from the three beams of the PAF after aligning the
data in sky coordinates. The inset shows the data from the three beams in the
same off-source region as in Fig.~\ref{fig13xy}a. As can be seen, the noise is
no longer correlated (i.e. $\rho_{m,n} = 0$ for $m \neq n$) and the RMS in the
averaged data from the three beams
reduces by $1/\sqrt{3}$ as expected for an uncorrelated noise.
\label{fig13xy}
}
\end{figure}

Survey speed is determined by the time required to integrate an image such that the sensitivity in each pixel is higher than some desired sensitivity. 
We consider surveying a region of the sky of angular size $\Omega_s$ sampled at $\Omega_b$, the independently sampled beam area in \eqref{eq:nb}. The number of pointings that need to be made for such an observation is $\Omega_s/\Omega_b$. 
Imaging speed can be improved with a multi-beam system by moving the telescope so that the region of interest is observed in each beam and then averaging the images obtained from different beams. As discussed above, for observing cases 1 and 2, the noise correlation between the beams does not affect the net sensitivity of the averaged image. However, the sensitivity of the beams are different, which need to be taken into account in the survey speed calculation. The effective number of beams for such applications is defined as 
\be
N_{\rm effb} = \frac{1}{\left(\frac{A\; \eta}{T_{{\rm sys}}}\right)_{\rm max}^2} \sum_{n=1}^{N_{\rm beam}} \left( \frac{A\; \eta_n}{T_{{\rm sys},n}} \right)^2
\label{eq:neffb}
\ee
where $A$ is the physical area of the telescope, $\frac{A\; \eta_n}{T_{{\rm sys},n}}$ is the
ratio of the effective area of the telescope to the system temperature for beam $n$ and
$\left(\frac{A\; \eta}{T_{{\rm sys}}}\right)_{\rm max}$ is the maximum value
of $\frac{A\; \eta_n}{T_{{\rm sys},n}}$. 
The survey speed figure of merit is
\be
{\rm SSFoM} \approx N_{\rm effb}\; \left(\frac{A\; \eta}{T_{{\rm sys}}}\right)_{\rm max}^2 
   \left(\theta_b \right)^2
\label{effssfom}
\ee
where $\theta_b$ is the square root of the independent beam area, which from the treatment of \cite{{haybird2015}} is $\lambda/(2D)$.  

\begin{figure}[!t]
\plotone{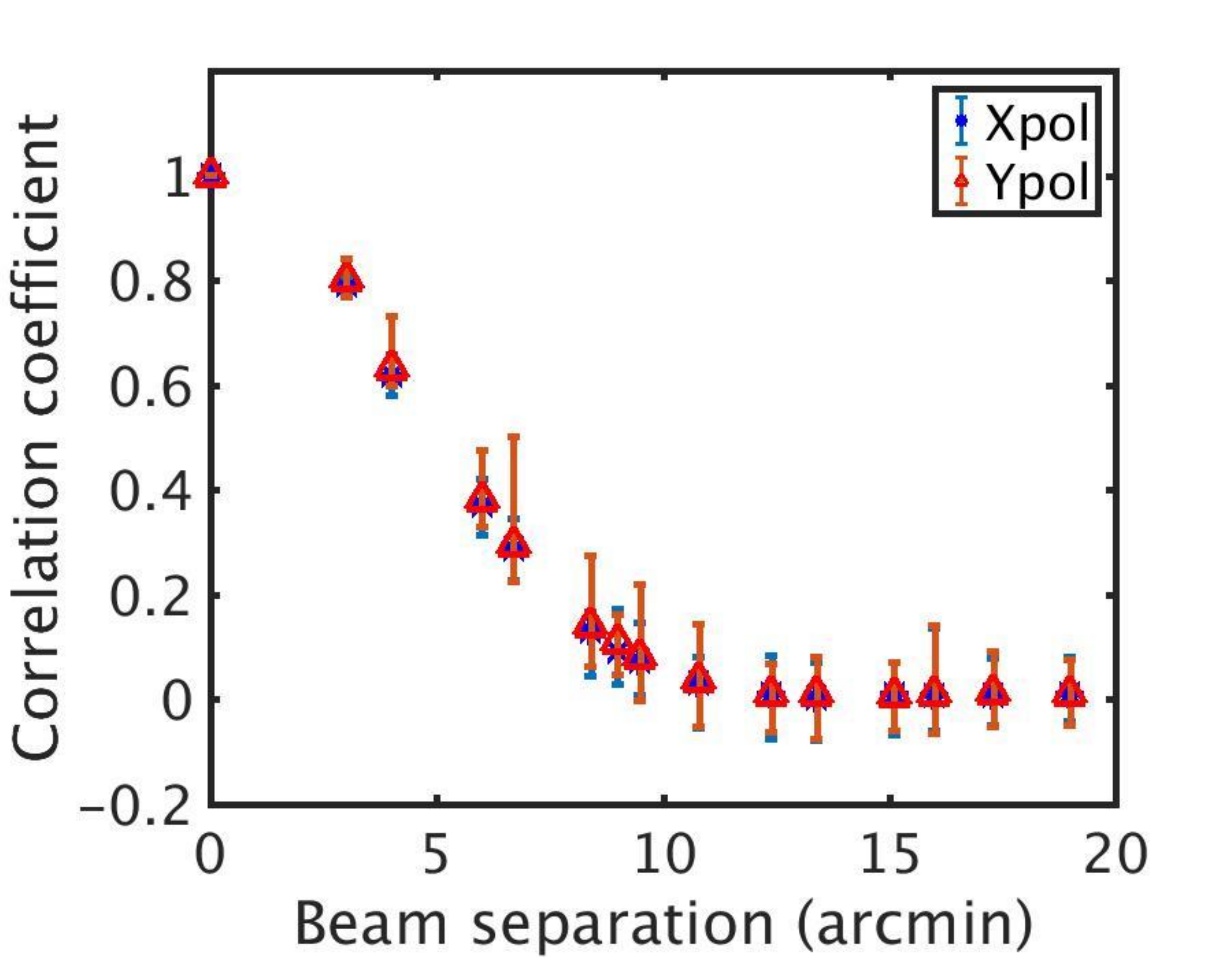}
\caption{Calculated noise correlation between beams of FLAG as a function of beam separation. The
`error bar' represents the peak-to-peak range in the calculated values.
\label{fig13xx}
}
\end{figure}


For observing case 3, we must first determine the beam to beam noise correlation, as represented in the fluctuations in the estimated power at the output of a formed PAF beams. We estimate the correlation coefficient as
\be
\rho_{n,m} = \frac{<(P_{B_n} - \overline{P}_{B_n})(P_{B_m} - \overline{P}_{B_m})>}
         {\sqrt{<(P_{B_n} - \overline{P}_{B_n})^2><(P_{B_m} - \overline{P}_{B_m})^2>}}.
\label{eqrho}
\ee
Here
\bea
P_{B_n} = P_{B_n}[j] & = \bm w_n^H \bm R[j] \bm w_n, \\
P_{B_m} = P_{B_m}[j] & = \bm w_m^H \bm R[j] \bm w_m
\eea
are the time series of the power from beam $n$ and $m$, respectively, estimated
from the time series of the correlation matrices $\bm R[j]$ after multiplying it
with the beamformer weights $\bm w_n$ and $\bm w_m$
for the two beams.  The expectations of $P_{B_n}$ and $P_{B_m}$ are obtained
as
\be
\overline{P}_{B_n}  = \frac{1}{M} \sum_{j=0}^{M-1} P_{B_n}[j], \\
\ee
\be
\overline{P}_{B_m}  = \frac{1}{M} \sum_{j=0}^{M-1} P_{B_m}[j].
\ee
The angle brackets in \eqref{eqrho} indicate time average over $M$ samples. A plot of the
calculated noise correlation between the boresight beam and the off-boresight beams
of FLAG as a function of beam separation is shown in Fig.~\ref{fig13xx}. The noise correlation
is obtained from the off-source data. The correlation drops
by about 60\% at the Nyquist beam separation ($\sim$ 4\arcmin) and it drops
to $\sim$ 15\% at twice the Nyquist beam separation.

The beam to beam noise correlation will result in a spatial correlation of noise in the image
made with a PAF. For a given beam spacing, the spatial correlation
coefficient of the noise in the image will be same as that given by \eqref{eqrho} even
though the correlation coefficient is obtained by
time average (the underlying stochastic process is ergodic in nature).
A lower angular resolution image needs to be made for the third class of applications,
but the S/N of the image will not improve by the square root of the number of pixels averaged
during smoothing. The improvement
in S/N depends on the angular resolution to which the image is smoothed. For the specific
case when the image is smoothed to an angular resolution approximately equal to SSFoV (see \eqref{eqssfov}), 
the final noise variance is 
\begin{equation}
  \sigma^2 = \frac{1}{N_\text{beam}^2} \sum_{m,n}^{N_\text{beam}} \sigma_m \sigma_n \rho_{n,m}
\end{equation}
where $\sigma_m, \sigma_n$ are the RMS noise fluctuations in beams $m$ and $n$, respectively, $\rho_{n,m}$ is the correlation coefficient of the noise in beams $m$ and $n$ (see \eqref{eqrho}) for the beam spacing used for the survey. The survey speed is inversely proportional to this noise variance. Approximating the sensitivity of the beams as equal and factoring the resulting survey speed as in \eqref{eq:nb} leads to the effective number of beams
\begin{equation}
  N_\text{effb} = N_\text{beam} \left(\frac{1}{N_\text{beam}} \sum_{m,n}^{N_\text{beam}} \rho_{n,m} \right)^{-1}
\end{equation}
If the correlation $\rho_{n,m}$ vanishes for $n \neq m$ (i.e., the noise is uncorrelated between beams), the number of effective beams is equal to the number of formed beams, as expected. 
Thus, the third class of observing application will benefit by keeping the
beams at twice the Nyquist separation or more since $\rho_{n,m}$ is smaller
for larger beam spacing (see Fig.~\ref{fig13xx}).

\subsubsection{Survey speed comparisons}

\begin{deluxetable*}{lrrrcr}[b!]
\tablecaption{Survey speed of FLAG\tablenotemark{a} \label{tabss}}
\tablecolumns{6}
\tablewidth{0pt}
\tablehead{
\colhead{Receiver} & \colhead{$\theta_b$\tablenotemark{b}} & \colhead{$T_{\rm sys}/\eta$\tablenotemark{c}} & \colhead{$N_{\rm effb}$\tablenotemark{d}} &
\colhead{SSFoM} & \colhead{Notes/Ref}  \\
\colhead{System} & \colhead{($\arcmin$)} & \colhead{(K)} & \colhead{} &
\colhead{(deg$^2$ m$^4$ K$^{-2}$)} & \colhead{}
}
\startdata
FLAG (Nyquist)    & 4  & 25.4, 25.7 & 6.9   & 2924  & 1,2   \\
FLAG (2 x Nyquist)  & 4  & 25.4, 30.5 & 5.2   & 2195  & 1,3   \\
FLAG (HI, Nyquist) & 3.8  & 28.0, 28.3 & 6.9   & 2184  & 1,4   \\
FLAG (HI, 2 x Nyquist)& 3.8  & 28.0, 33.5 & 5.2   & 1640  & 1,5   \\
\enddata
\tablenotetext{a}{The physical area of the GBT aperture is taken as 7854 m$^2$ for the calculation of SSFoM using \eqref{effssfom}.}
\tablenotetext{b}{$\theta_b = \frac{\lambda}{2\;D}$ is taken as the Nyquist beam separation. }
\tablenotetext{c}{$T_{\rm sys}/\eta$ for the central beam and the 6 outer beams.}
\tablenotetext{d}{$N_{\rm effb}$ is computed using \eqref{eq:neffb}.}
\tablecomments{1. Ref. This paper \\
2. Spectroscopic observations near 1350 MHz with Nyquist  ($\sim$4\arcmin) beam separation.  \\
3. Spectroscopic observations near 1350 MHz with twice Nyquist  ($\sim$8\arcmin) beam separation.  \\
4. \HI\ observations near 1420 MHz with Nyquist  ($\sim$4\arcmin) beam separation.  \\
5. \HI\ observations near 1420 MHz with twice Nyquist  ($\sim$8\arcmin) beam separation.
}
\end{deluxetable*}

\begin{deluxetable*}{lrrrrrcccr}[b!]
\tablecaption{Survey speed comparison \label{tabss1}}
\tablecolumns{10}
\tablewidth{0pt}
\tablehead{
\colhead{Receiver} & \colhead{A} & \colhead{Freq} & \colhead{$\theta_b$\tablenotemark{a}} &
\colhead{$T_{\rm sys}/\eta$\tablenotemark{b}} & \colhead{$N_{\rm effb}$} 
& \colhead{SSFoM\tablenotemark{c}} & 
\colhead{BW\tablenotemark{d}} & \colhead{SSFoM$\times$BW} & \colhead{Notes/Ref}  \\
\colhead{System} & \colhead{(m$^2$)} & \colhead{(GHz)} & \colhead{($\arcmin$)} &
\colhead{(K)} & \colhead{} & \colhead{(deg$^2$ m$^4$ K$^{-2}$)} & 
\colhead{(MHz)} & \colhead{(deg$^2$ m$^4$ K$^{-2}$ MHz)} & \colhead{}
}
\startdata
FLAG (Nyquist)    & 7854  & 1.35 & 4  & 25.4, 25.7 & 6.9   & 2924  & 150 & 3.6 $\times$ 10$^5$ & 1,7   \\
FLAG     & 7854  & 1.35 & 4  & 25.4 & $\sim$ 25   & 10690 & 150 & 1.6 $\times$ 10$^6$ & 1,8   \\
GBT L-band        & 7854  & 1.35 & 4 & 25.7 & 1.0   & 415   & 400 & 1.7 $\times$ 10$^5$ & 2  \\
Parkes multi-beam & 3217  & 1.37 & 5.9 & 41.9 & 13  & 1010   & 300 & 3.0 $\times$ 10$^5$ &  3,9  \\
Arecibo ALFA      & 35633 & 1.37 & 1.8 &  35.2, 45.5 & 4.6  & 4228  & 300 & 1.3 $\times$ 10$^6$ & 4,10  \\
Effelsberg 7beam  & 7854  & 1.41 & 3.7 & 45.8 &  7 & 782   & 300 & 2.3 $\times$ 10$^5$ & 5,11  \\
Parkes with PAF   & 3217  & 1.31   & 5.9  & 60.0 & $\sim$ 144  & 4025  & 700 & 2.8$\times$ 10$^6$ & 6  \\
\enddata
\tablenotetext{a}{$\theta_b = \frac{\lambda}{2\;D}$ is taken as the Nyquist beam separation for all telescopes.}
\tablenotetext{b}{$T_{\rm sys}/\eta$ are provided for the central beam and for all the outer beams for cases where two values are listed.}
\tablenotetext{c}{Survey metric for spectroscopic observations.}
\tablenotetext{d}{Real time signal processing bandwidth.}
\tablecomments{1. Ref. This paper \\
2. Ref. \url{https://science.nrao.edu/facilities/gbt/proposing/GBTpg.pdf} \\
3. Ref. \url{http://www.parkes.atnf.csiro.au/observing/documentation/user_guide/pks_ug_3.html} \\
4. Ref. \url{http://www.naic.edu/alfa/gen_info/info_obs.shtml} \\
5. Ref. \url{https://eff100mwiki.mpifr-bonn.mpg.de/doku.php?id=information_for_astronomers:rx:p217mm} \\
6. Ref. \citet{chippendaleetal2016}. SSFoM at 1.31 GHz is obtained as 
SSFoV $\times$ $\mathcal{S}_{\rm max}^2$, where SSFoV=1.4 deg$^2$ and 
$\mathcal{S}_{\rm max} = 3217/60 = 53.6$ m$^2$ K$^{-1}$. $N_{\rm effb} \sim \frac{{\rm SSFoV}}{\theta_b^2} = 144$ . The real-time processing bandwidth is assumed to be same as the front-end bandwidth of $\sim$ 700 MHz.\\
7. FLAG with 7 formed beams placed at Nyquist  ($\sim$4\arcmin) beam separation.  \\
8. Intrinsic SSFoM (see \eqref{eq:ssfom_hay}) of FLAG obtained using the data shown in Fig.~\ref{fig7}a. SSFoV=0.11 deg$^2$; $N_{\rm effb} \sim \frac{{\rm SSFoV}}{\theta_b^2} = 25$. \\
9. 1.1~Jy~K$^{-1}$ is used for the calculation of SSFoM. \\
10. The region illuminated by ALFA is taken as $\sim$ 213 m in diameter.  11~K~Jy$^{-1}$ for the central beam and 8.5~K~Jy$^{-1}$ for outer beams are used for the calculation of SSFoM. \\
11. Aperture efficiency of 48\% used for the calculation of SSFoM.
}
\end{deluxetable*}

The SSFoMs of FLAG, obtained using \eqref{effssfom}, with 7 beams spaced 
at Nyquist separation ($\sim$ 4\arcmin) and
twice Nyquist separation ($\sim$ 8\arcmin) are listed in Table~\ref{tabss}.
A major use of the new PAF system on the GBT will be to observe
extended \HI\ emission from nearby galaxies. These observations
will be typically done in the OTF mode and fall into the first and second category
of observations mentioned above.  The $T_{\rm sys}/\eta$
of FLAG is optimum near 1350 MHz and it degrades by a factor of $\sim$ 1.1
near 1.42 GHz. Thus the SSFoMs for \HI\ observations 
are 2184 and 1640 deg$^2$ m$^4$ K$^{-2}$ for the
two beam separations (see Table~\ref{tabss}), which are a factor of 5.3 and 4 higher than the
GBT single-feed receiver.
A comparison of the SSFoM of FLAG, 
the existing GBT single-feed receiver, and our estimated values for
multi-beam receivers at other telescopes, is given in Table~\ref{tabss1}.
We computed the intrinsic SSFoM (see \eqref{eq:ssfom_hay}) of FLAG 
using the data shown in Fig.~\ref{fig7}a, which is about 10690 deg$^2$ m$^4$ K$^{-2}$.
The derived SSFoV using \eqref{eqssfov} shows that about 25 beams 
need to be formed in the signal processing backend to obtain the
full survey capability of FLAG (see Table~\ref{tabss1}). 

For broadband observations (for example, pulsar observations), the survey capability of an instrument scales with the signal processing bandwidth. The real-time beamformer backend for FLAG will have a bandwidth of about 150~MHz.
The mean value of $T_{\rm sys}/\eta$ for the 7 beams over this bandwidth is about
28~K for Nyquist beam separation.
The SSFoM scaled by bandwidth is 3.6 $\times$ $10^5$~deg$^2$~m$^4$~K$^{-2}$~MHz.
For the GBT single feed L-band system the usable bandwidth is
about 400~MHz (after removing the RFI effected bands). Assuming a
constant $T_{\rm sys}/\eta$ over the 400~MHz bandwidth of the GBT, the
survey speed of the single feed system is 1.7 $\times$ $10^5$~deg$^2$~m$^4$~K$^{-2}$~MHz.
The SSFoM bandwidth product of FLAG is about 2.1 times
larger than the GBT single feed system for pulsar or other broadband
applications.
}

\subsection{Comparison with the PAF model, system parameters and FoV}
\label{compare}

\begin{figure}[!t]
\plottwo{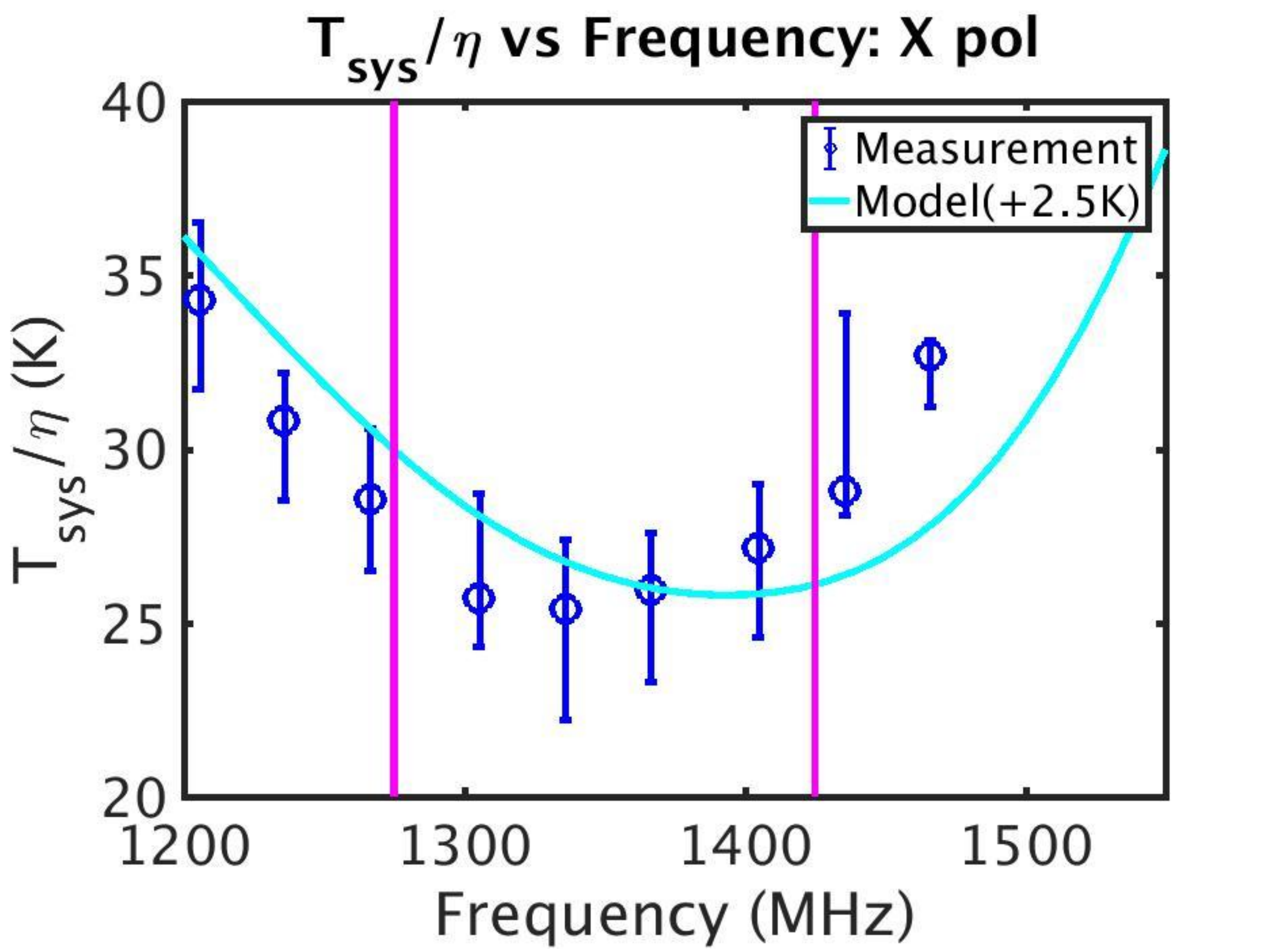}{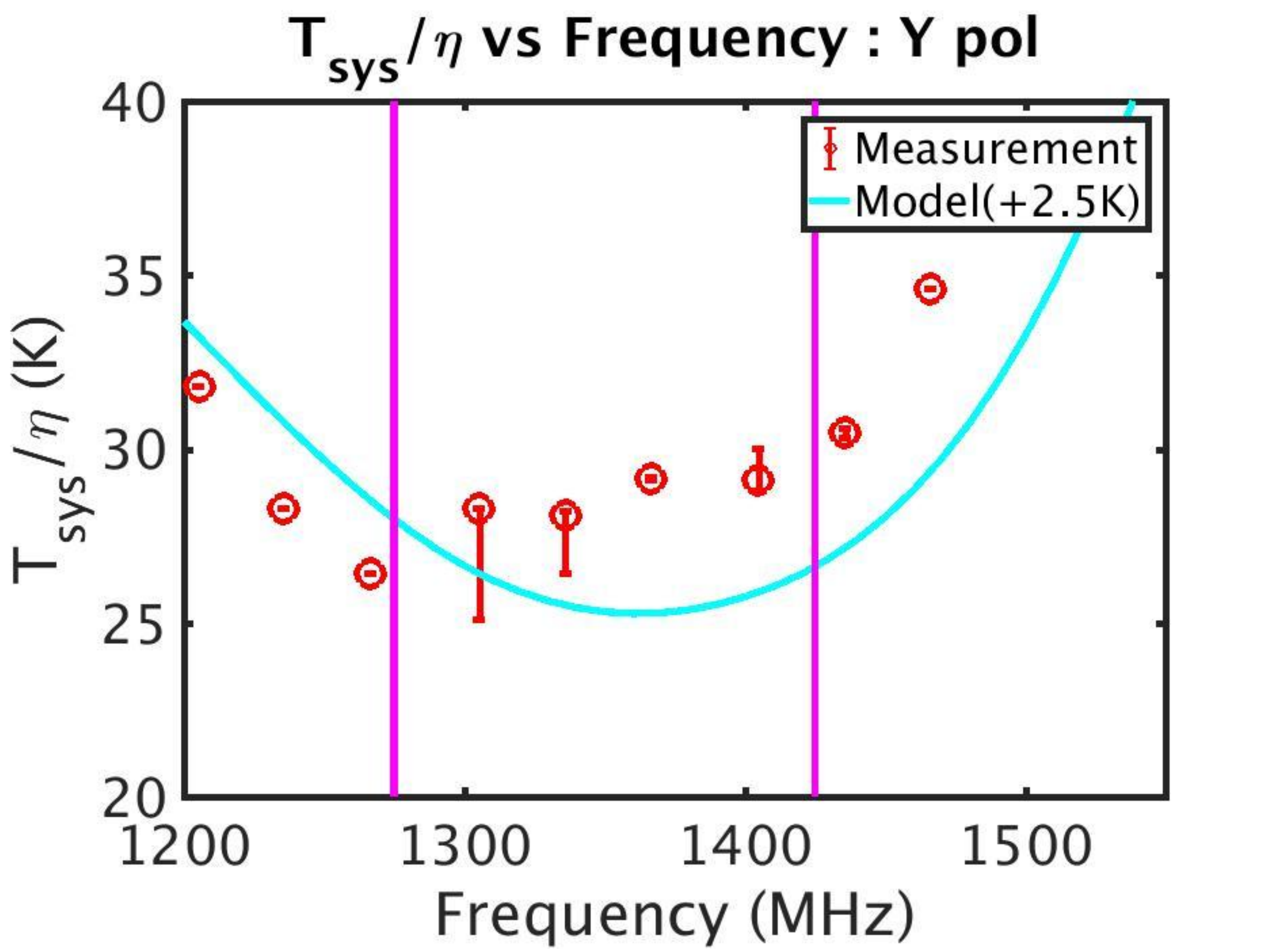}%
\caption{
The PAF model prediction (solid line) along with the median measured
$T_{\rm sys}/\eta$ with their peak-to-peak
variation for X (left) and Y (right) polarizations. The data points shown in
Fig.~\ref{fig10} are used to compute the median $T_{\rm sys}/\eta$. The median is
computed from the set of measurements in a frequency interval of
$\sim 1.5$~MHz. The model assumes loss-less PAF and hence the system temperature in the
model is increased by 2.5~K to take into account of the losses ahead
of the LNA (see Fig.~\ref{fig3}b).
\label{fig13}
}
\end{figure}

\begin{figure}[!t]
\plottwo{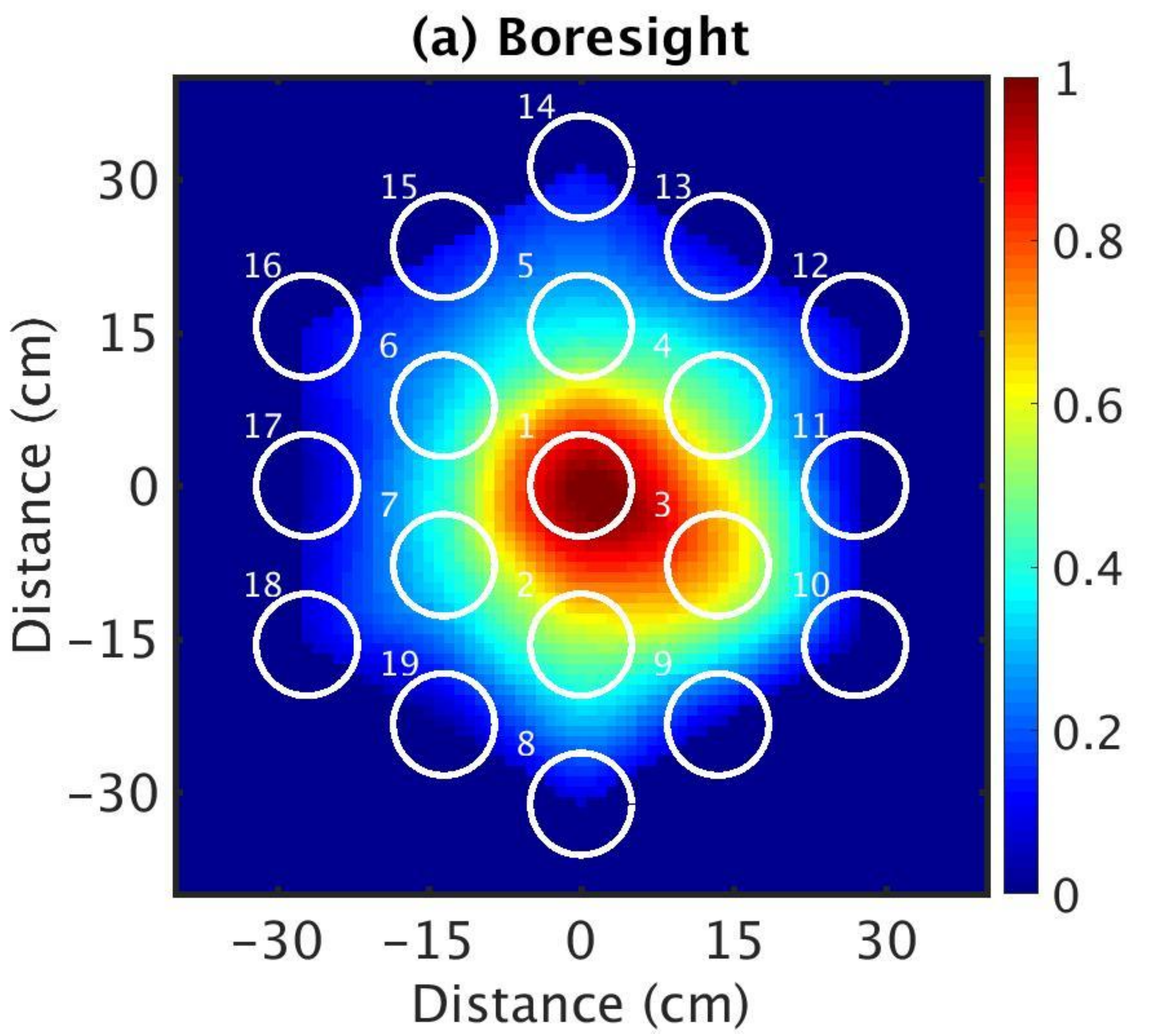}{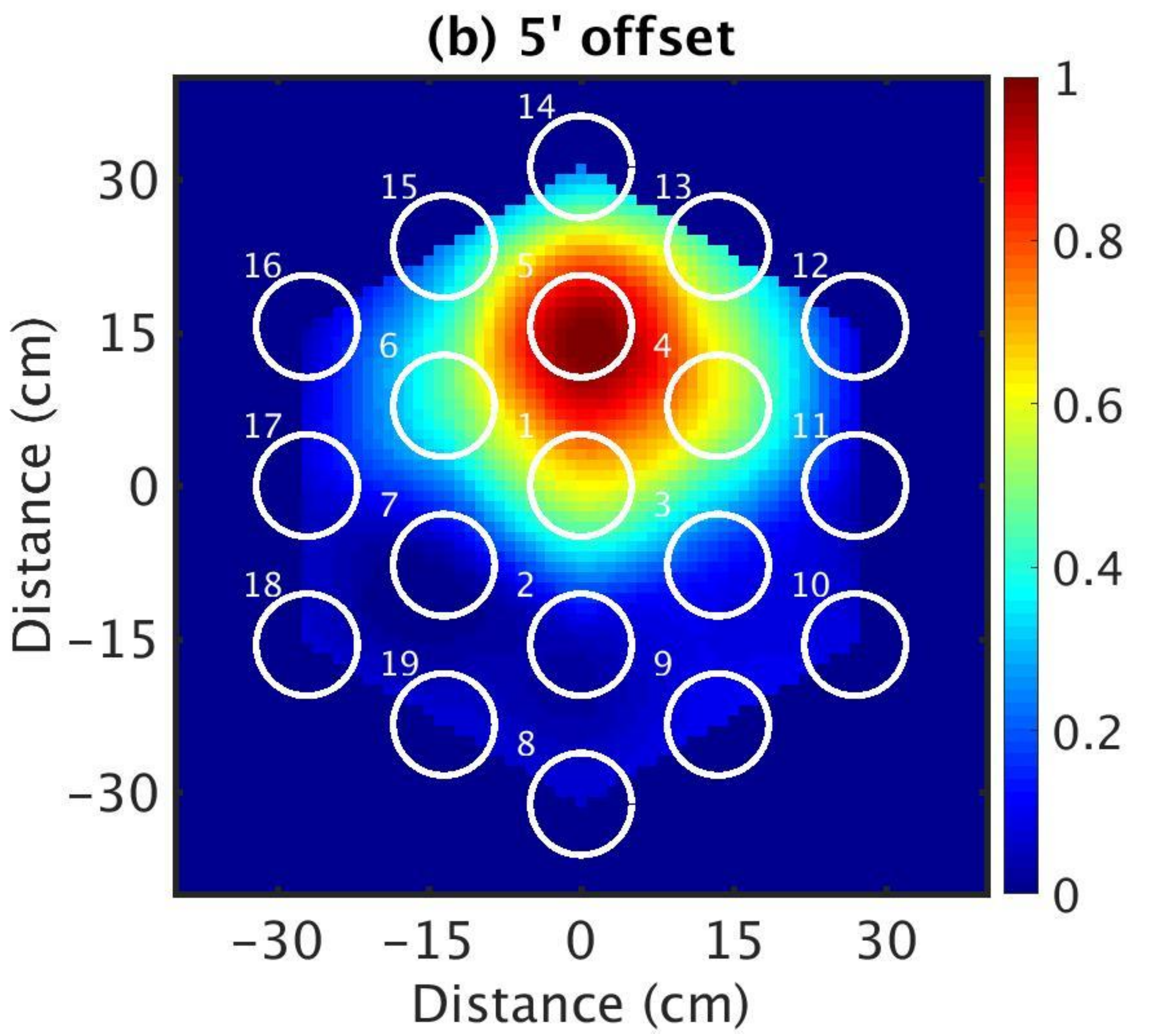}
\caption{(a) The normalized amplitude distribution of the weights
  obtained for the boresight direction superposed on the dipole array
  geometry. The white circles indicate the location of the dipoles in
  the array along with the dipole number. (b) Same as (a) but for
  5\arcmin~offset from boresight direction.
\label{fig14}
}
\end{figure}

\begin{deluxetable*}{lr}[b!]
\tablecaption{System parameters \label{tab3}}
\tablecolumns{2}
\tablewidth{0pt}
\tablehead{
\colhead{Parameter} & \colhead{Value}
}
\startdata
\twocolhead{Measured parameters} \\ \hline
$T_{\rm sys}/\eta$ near 1350~MHz & 25.4$\pm$2.5~K \\
LNA noise temperature &  5~K \\
Inferred loss ahead of LNA\tablenotemark{a} & 2.5~K \\
Cosmic Microwave Background temperature & 2.7~K \\
Galactic Background Temperature\tablenotemark{b} & 0.8~K \\\hline
\twocolhead{Estimated parameters} \\\hline
Atmospheric Temperature\tablenotemark{c} & 2~K \\\hline
\twocolhead{Inferred from Model} \\\hline
Beamformed Receiver temperature & 7.5~K\\
Spillover Temperature & 3.5~K \\
Total System Temperature & 16.5~K \\
Estimated Spillover Efficiency & 98.8\% \\
Estimated Aperture Efficiency & 60\% \\
Model $T_{\rm sys}/\eta$ near 1350~MHz & 27.5~K\tablenotemark{d} \\
\enddata
\tablenotetext{a}Model results with an additional noise of 2.5~K
to account for losses ahead of the LNA (see Fig.~\ref{fig3}b)
are in qualitative agreement with the measurements
for the amplifier noise parameters used here.
\tablenotetext{b}{Median of the off-source sky temperature estimated from the 1.4~GHz survey data of
\citet{reichreich1986}}
\tablenotetext{c}{\citet{jean-yvesetal2002}}
\tablenotetext{d}{The uncertainty in the model value is up to 20\% (see text)}
\end{deluxetable*}

For high-sensitivity receivers, further reductions in system noise
become increasingly challenging as the system performance
improves. This is especially true for phased array receivers, for
which mutual coupling effects require a holistic approach to the
design optimization of the array elements and front end
electronics. Extensive modeling efforts were critical to the PAF
design optimization and understanding the system performance
\citep{warnicketal2011, warnicketal2009, roshifisher2016}.  The steps
involved in the modeling are the following
\citep{roshifisher2016}. The dipole array was first modeled using a
full wave finite element solver in CST microwave studio\footnote{ a
  commercial 3D electromagnetic simulation software;
  \url{https://www.cst.com/}} to obtain the element beam patterns and
the impedance matrix.  The embedded beam patterns are then obtained
from the element patterns.  The secondary radiation patterns for the
GBT optical geometry were obtained using a physical optics
approximation. The embedded patterns along with the GBT geometry were
used to compute the noise covariance matrices due to ground
spillover. The secondary patterns were used to compute the signal
response due to the source and noise covariances due to the sky
background radiation.  The impedance matrix of the array combined with
a noise model for the cryogenic LNAs \citep{pospieszalski2010}
provided the receiver noise covariance. The amplifier noise parameters
used for the modeling are: minimum receiver temperature $T_{\rm min}$
= 4.2~K, the optimum impedance $Z_{opt}$ = $28.9-j3.5$ $\Omega$, and
Lange parameter $N = 0.007$. The noise parameters are considered to be
approximately constant over the frequency range 1200 to 1550~MHz. This
LNA noise model is obtained from the amplifier modeling and reproduces
the measured LNA noise temperature, however, we note that these
parameters cannot be uniquely constrained from noise temperature
measurements alone. Accurate noise modeling of the LNA and the measurement
of noise parameters are underway -- PAF model results with these
new values will be presented elsewhere.
The signal response and noise covariances were
used to compute the expected S/N. The maximum S/N beamformer algorithm
was then used to find beamformer coefficients for each desired beam
steering direction. The model was run repeatedly for a set of
frequencies ranging from 1200 to 1550~MHz. With the accurate
representation of the PAF by input parameters, the model can predict
the receiver temperature, antenna temperature, spillover temperature,
and the full polarization electromagnetic fields in the antenna aperture. Below we
compare the measured system performance with PAF model predictions.

The PAF model prediction for the boresight direction is shown in
Fig.~\ref{fig13} along with the measured median $T_{\rm sys}/\eta$ as
a function of frequency
\footnote{The model curve is not a least square fit to the
  data. Currently, the major contribution to the uncertainty in the modeled value is the
  inaccuracies in one of the input parameters to the model, the amplifier noise parameters and their frequency
  dependence. We are in the process of accurately modeling the
  amplifier noise parameters and measuring them. The PAF model results
  based on these new measurements will be presented elsewhere.}  The
median values were computed from the measured data points shown in
Fig.~\ref{fig10} over a frequency interval of $\sim$ 1.5~MHz. The
median values for Y-polarization were computed from the subset of
measured values that is not severely affected by the pointing offset
and faulty dipoles (see Section~\ref{boresight}). The model results
are plotted with an additional noise contribution to $T_{\rm sys}$
to account for the losses ahead of the LNA. For the LNA noise parameters
used here, model results with an additional noise of 2.5~K (see Fig.~\ref{fig3}b)
are in qualitative agreement with the measurements at frequencies
below 1.45~GHz.
The discrepancy between model and measurement at frequency above 1.45~GHz
may be due to a combination of the following factors: inaccuracy in
the amplifier noise parameters, error in the electromagnetic
simulation, unmodeled ground scattering due to the feed support
structure, or manufacturing errors in the dipoles. Ground scattering
from the feed support may have additional contributions at higher
frequencies where the increase in system temperature is dominated by
the presence of array grating lobes. These possibilities are the
subject of an ongoing investigation.

The model prediction for $T_{\rm sys}/\eta$ versus offset from the
boresight is shown in the Fig.~\ref{fig8}. The model results, obtained
with the 2.5~K excess noise due to the losses upstream of the LNA
(see Fig.~\ref{fig3}b), tracks closely with the measured
variation of $T_{\rm sys}/\eta$ as a function of the radial offset
very well.  The increase in $T_{\rm sys}/\eta$ for offsets larger than
5\arcmin\ is due to the finite size of the dipole array. This is
evident from Fig.~\ref{fig14}, where we plot the normalized beamformer
weight distribution over the dipole array geometry for the boresight
beam, and a beam $\sim$5\arcmin\ offset from the boresight direction.
These weight distributions are obtained from grid observation data
toward Virgo A.  As seen in the figure, significant amplitudes for the
weights are clustered around seven dipole elements, roughly following the
Airy pattern due to the compact source. At offsets \gsim~5\arcmin\ the
cluster of seven elements is located at the edge of the array centered on
dipole 5 (see Fig.~\ref{fig14}b).  Thus at offsets more than
5\arcmin~from the boresight the dipole array does not have enough
elements to sample the Airy pattern well. Thus the FoV limitation of the
array indicated by the upward slope of $T_{\rm sys}/\eta$ in
Fig.~\ref{fig8} is caused only by the limited extent of the array, and
thus could be extended by the addition of more elements.

On-telescope measurements do not provide $T_{\rm sys}$ and $\eta$
separately and so we infer these values and other system parameters
from the PAF model.  The inferred system parameters are summarized in
Table~\ref{tab3}. The system temperature after forming the beams is
about 16~K, with contributions from receiver noise of 7.5~K, spillover
of 3.5~K and sky background plus atmosphere of 5.5~K. The median
increase in formed beam S/N on a compact source is about a factor of
eight compared to a single dipole near 1336~MHz. This increase in S/N
implies a $T_{\rm sys}/\eta$ of $\sim$ 216~K for the single dipole
case - based on scaling the measured $T_{\rm sys}/\eta$ for the formed
beam (see Table~\ref{tab3}). This high $T_{\rm sys}/\eta$ is due to
the large spillover contribution when observing with a single dipole.
The inferred spillover temperature from the model is $\sim$100~K for
the single dipole case.  The spillover efficiency is increased to
about 98\% in the process of beamforming, thus reducing the modeled
$T_{\rm sys}/\eta$ to about 27~K for the formed beam. The high ground
suppression achieved for the formed beam, results in a somewhat lower
aperture efficiency (about 60 \%) compared to the GBT single feed
system, an inevitable trade off for the 19-element prime focus PAF.

The uncertainty in the model predictions have several contributions,
which include: (a) amplifier noise parameters and their frequency
dependence; (b) accuracy of the CST simulation results and (c) model
not accounting for the scattering due to feed support structures.  The
estimated values of $T_{\rm sys}/\eta$ at frequencies below $\sim$
1300~MHz sensitively depends on the amplifier noise parameters due to
a higher level of mutual coupling in the array. At frequencies above
$\sim$ 1450~MHz the grating lobes become significant and hence the
scattering due to the feed support structure could limit the accuracy
of the computation.  In the frequency range near 1350~MHz we estimate
that the model predictions are accurate to within 20\%. This estimation
of accuracy is obtained by considering different amplifier noise
parameters that are consistent with the LNA noise temperature measurements
and examining the variation of the computed $T_{\rm sys}/\eta$ near 1350~MHz.

\section{Observations of Astronomical sources}
\label{astron}

\begin{figure}[!t]
\plottwo{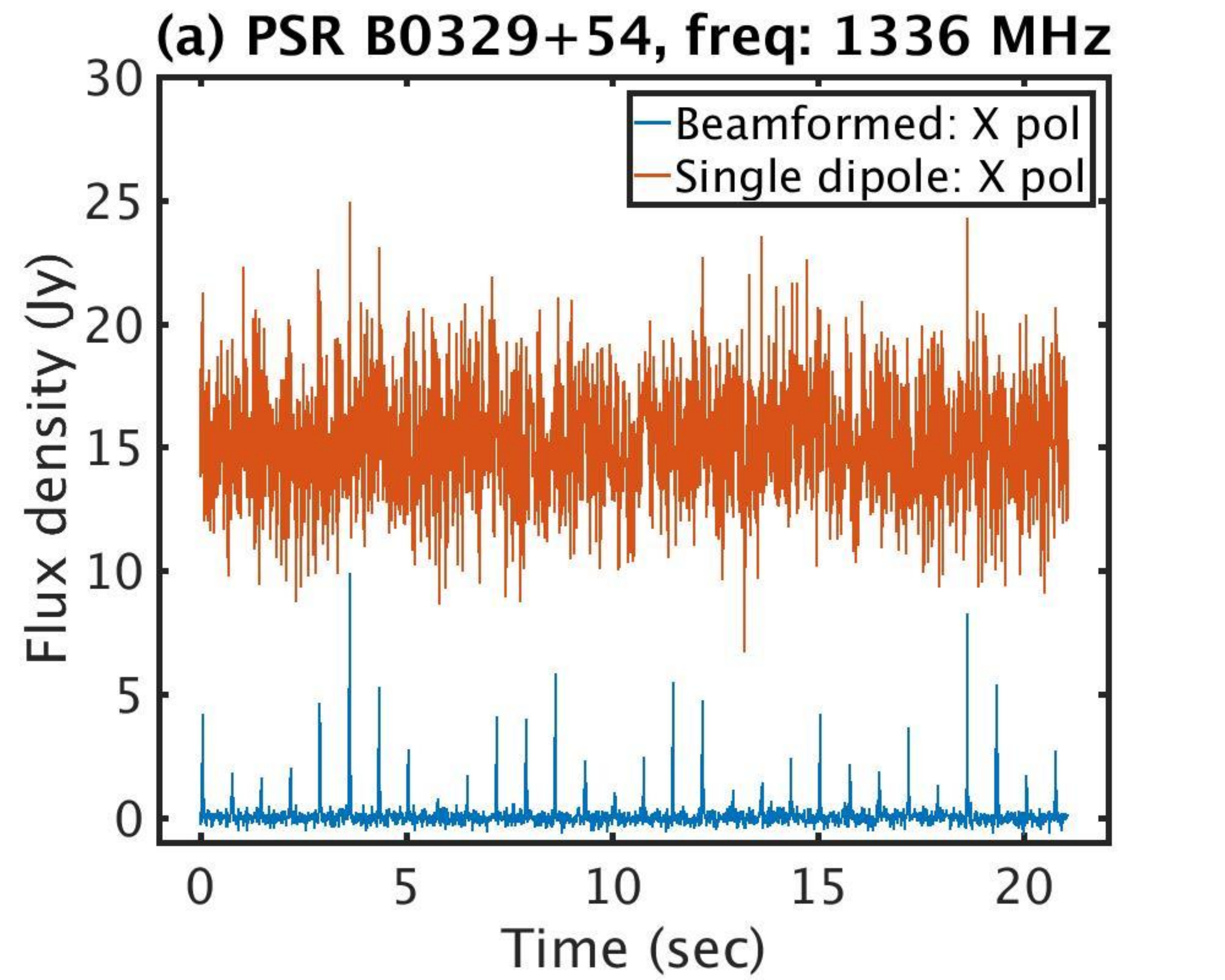}{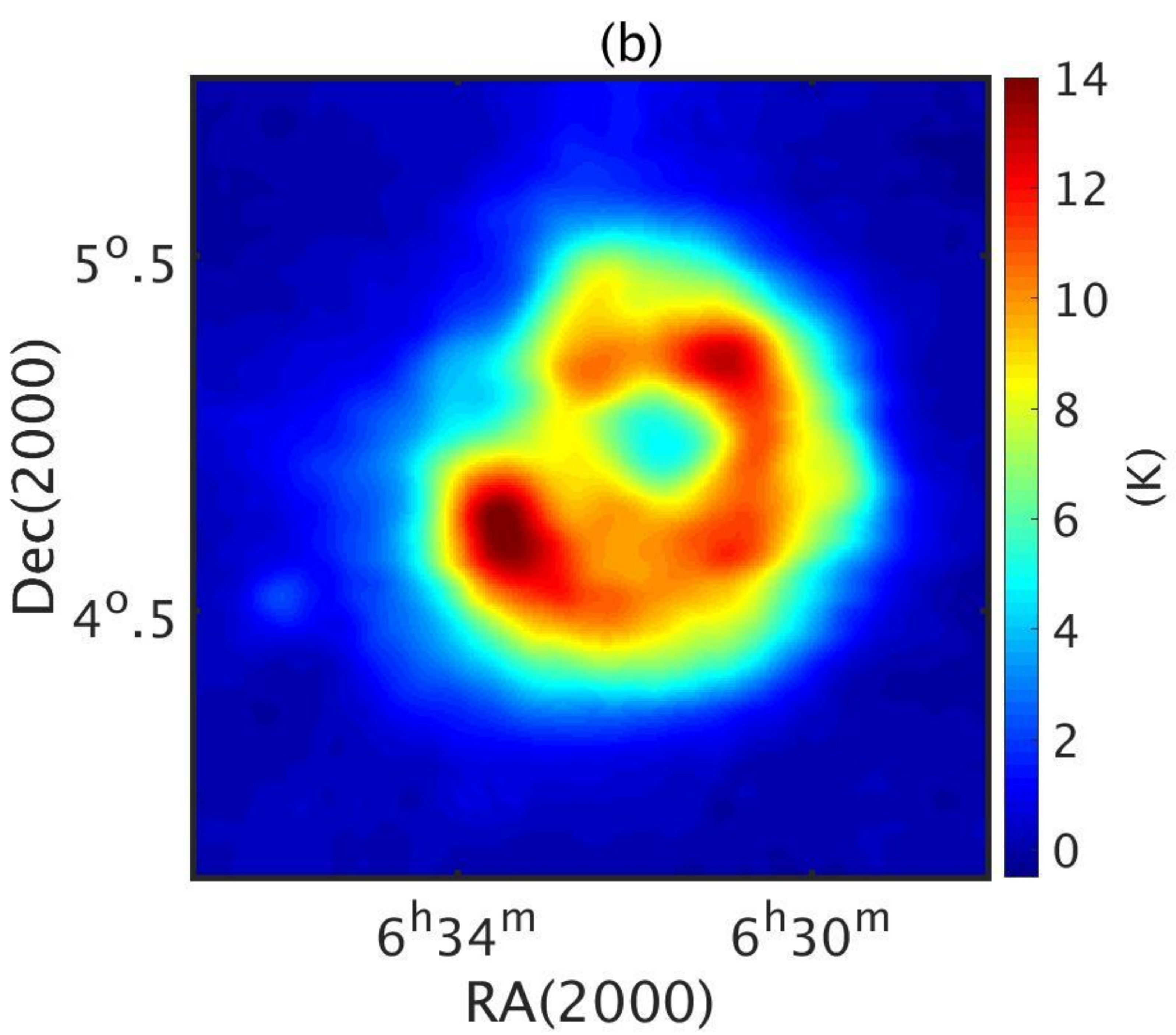}
\caption{(a) Observation of the pulsar B0329+54 with a single dipole
  (top) and with maximum S/N boresight beam. The increase in S/N is
  about a factor of eight. (b) Image of the Rosette nebula made with
  the PAF near 1.336~GHz.
\label{fig15}
}
\end{figure}

We have observed the pulsar B0329+54 and the Rosette Nebula
with the PAF system on the GBT.
The data taken towards these sources were obtained using the
experimental setup described in Section~\ref{secinst} and processed
as described in the Section~\ref{secpm_dp}.

\subsection{PSR~B0329$+$54}

The pulsar B0329+54 was observed with the PAF receiver on March 16,
2017.  The observed frequency was 1336.0275~MHz with a bandwidth of
~300 kHz. The pulse width of B0329+54 at 10\% of the peak average
pulsar amplitude is 31.4~ms \citep{manchesteretal2005}.  Therefore,
the cross correlations were integrated for about 10~ms. An on-off
observation on the calibrator 3C123 was performed before taking the
pulsar data in the same observing setup. The data set on the
calibrator was used to obtain the beamformer weights. A time series
from the pulsar data was then obtained by estimating the power using
the beamformer weights for every 10~ms. This power was converted
into flux density units using the calibration factor derived from the
3C123 observations.  The calibrated, beamformed time series from X
polarization data is shown in Fig.~\ref{fig15}a (bottom curve). For
comparison the time series obtained from the central dipole is also
shown in Fig.~\ref{fig15}a (top curve). The improvement in S/N in the
formed beam output is about a factor 8, similar to what is measured
from observations of calibrator sources. This indicates that the
transfer of beamformer weights from calibrator observation gives the
expected improvement in S/N on the target source.

\subsection{Rosette Nebula}

The continuum emission from Rosette Nebula was observed at frequency
1336.0275~MHz with a bandwidth of $\sim$ 300 kHz. The telescope was
moved in a raster scan mode along right accession  and declination
while recording the voltages. The data from each row of the
raster scan were recorded to a file and processed offline as described
in Section~\ref{secpm_dp}. The integration time for the cross
correlations was set to $\sim 170$~ms. The telescope speed for the
raster scan was such that it moved by about 1\arcmin\ (1/10$^{\rm th}$ of the
beamwidth) in the sky during this integration time. The time stamp on
the data and that on the telescope position were used to obtain the
sky position corresponding to each integration time.

The synchronization between the data acquisition system and the
telescope for the raster scan mode of observing was not robust. This
synchronization issue had two effects: (a) the sky position derived
had to be corrected manually to get the true equatorial coordinates of the
observed positions; (b) we lost data for a few right ascension scans,
which corresponded to a gap of about 28\arcmin\ in declination in the
image. As described below, this gap is filled with data obtained from
different beams of the PAF.

We observed the calibrator 3C123 along with Rosette observations in
order to derive the beamformer weights. However, this data set could
be used only to obtain the weights for the boresight beam and another
off-boresight beam (5\arcmin\ in elevation toward north) due to a
telescope pointing offset. The images made from these two beams had
the expected sensitivity. But due to the loss of data, the image had
gaps and could not be filled with the data from two beams alone.  We
therefore derived the beamformer weights from grid observation toward
Virgo A taken 2 days before the Rosette observations. These beamformer
weights did not provide the optimum signal to noise ratio due to the
temporal change in instrumental gain and phase between the Virgo A and
Rosette observations. The degradation in S/N was
about 20\%.  The Virgo A data set was used to form images from
different beams and to calibrate the estimated power in Jy. The PAF
system did not have a calibrated noise source ahead of the LNA and
hence converting the flux density scale to brightness temperature
scale had some uncertainty. Further, the data from beams outside the
nominal FoV of the PAF ($\sim$ 20\arcmin) were used to fill the gaps
in the image (see above).  The calibration factor to convert Jy to
brightness temperature in~K had to be increased by 33\% for these
beams.  Thus, the overall accuracy of the brightness temperature scale
of the image is estimated to be about 30 \%. Determining the Jy to K
conversion and its stability for the PAF system are part of ongoing
research work.

After calibration, a linear baseline, estimated using the
data points away from the Rosette nebula, was subtracted from each
right ascension scan. The variation of the mean value of the baseline from scan to scan was $\sim$
10\%, which may be due to system gain variation.  The image obtained
from the baseline subtracted data and after combining the data from 3
beams (central beam, a beam 9\arcmin\ north in elevation and a beam 9\arcmin\
south in elevation) is shown in Fig.~\ref{fig15}b. The combined
image is smoothed with a box function of $\sim$ 8\arcmin\ $\times$
8\arcmin\ in size.

The features of the Rosette nebula seen in Fig.~\ref{fig15}b compare
well with those observed earlier by \cite{celnik1985} at 1410~MHz. The
manual adjustment of the sky position needed due to loss of
synchronization mentioned above had resulted in a residual artifact at
about 2~K level in the image. This artifact is entirely due to the
synchronization between data acquisition and telescope control systems
and is not due to the PAF.

\section{Summary and Conclusion}
\label{sumcon}

We presented the measured performance of FLAG front-end, a new
1.4~GHz 19-element, dual-polarization, cryogenic phased array feed
radio astronomy receiver built for the GBT. A brief description of the
instrumentation was given, which included a novel method of
implementing an unformatted digital link.  The performance of the
system was measured by placing the PAF at the prime focus of the GBT
and observing a set of astronomical calibrators. The performance
metric, $T_{\rm sys}/\eta$, had a median value of 25.4~K $\pm$ 2.5~K
near 1350~MHz.  This value is comparable to the performance of the
single feed system of the GBT at 1.4~GHz. The median $T_{\rm
  sys}/\eta$ was higher by about 5~K near the edge of the 150~MHz
bandwidth of interest, centered at 1350~MHz. The increase in $T_{\rm
  sys}/\eta$ at 1336~MHz at $\sim$ 4\arcmin\ offset, required for Nyquist
sampling, was $\sim 1$\% and at $\sim$ 8\arcmin\ offset was $\sim 20$\%.
The distribution of $T_{\rm sys}/\eta$
in elevation and cross-elevation directions was radially
symmetric. This symmetry enables the PAF to form seven high sensitivity
beams within the FoV, resulting in an increase in survey
speed by a factor between 2.1 and 7 depending on the observing
application. The FoV of the PAF system is limited by the size of the array,
as there are not enough elements to form a high sensitivity beam for
offset angles \gsim~5\arcmin. The PAF model predictions qualitatively
agree with the measured variation of $T_{\rm sys}/\eta$ with frequency
as well as with offset from boresight.
The results from the observations of a pulsar and an
extended source with the PAF system on the GBT are also presented.

The results presented here were processed by a narrow-band offline
processing system. Future observations will use a new real-time
150~MHz bandwidth digital signal processing system developed by
FLAG collaboration.  The PAF and the broadband beamformer comprise the
complete FLAG instrumentation, which will enable efficient searches
for pulsars and Fast Radio Bursts and observations of diffuse,
extended neutral hydrogen emission in the circumgalactic medium of
nearby galaxies.

\acknowledgments

The authors would like to thank Sivasankaran Srikanth, Marian
Pospieszalski, Anthony Kerr, Morgan Mcleod, Bob Dickman, S. K. Pan,
Brent Carlson, Bill Randolph, Mike Shannon, Gerry Petencin, Kamaljeet
Saini, Greg Morris and Nicole Thisdell of the National Radio Astronomy
Observatory, and Karen O'Neil, Joe Brandt, Dennis Egan, Mike Hedrick,
Pat Schaffner, Roger Dickenson, Bob Anderson, Ron Maddalena and Jonah
Bauserman of the Green Bank Observatory. Thanks to Lewis Ball and Brian
Mason of NRAO for review of a previous version of the text. We
thank the anonymous referee for the critical comments on the manuscript,
particularly on the noise correlation between the PAF beams and its effect on
the survey speed.

\end{document}